\documentclass[3p,times]{elsarticle}

\usepackage{hyperref}
\usepackage{color}

\usepackage[table,xcdraw]{xcolor}
\usepackage{ecrc}
\usepackage{amsmath}
\usepackage{csquotes}
\volume{00}

\firstpage{1}

\journalname{Computer Law and Security Review}

\runauth{Hauer \& Kevekordes et al.}

\jid{procs}

\jnltitlelogo{Procedia Computer Science}

\usepackage{amssymb}
\usepackage[figuresright]{rotating}

\begin{document}

\begin{frontmatter}



\dochead{This is a preprint! The final version hast been published by the Computer Law and Security Review journal (https://doi.org/10.1016/j.clsr.2021.105583)}

%\title{AI in HR - a legal perspective on possible fairness measures}
\title{Legal perspective on possible fairness measures - A legal discussion using the example of hiring decisions}

\author[marc]{Marc P. Hauer}
\ead{hauer@cs.uni-kl.de}

\author[johannes]{Johannes Kevekordes}
\ead{j\_keve01@wwu.de}

\author[maryam,maryam2]{Maryam Amir Haeri}
\ead{haeri@informatik.uni-kl.de}

\address[marc]{[OrcID 0000-0002-1598-1812], Algorithm Accountability Lab (AAL) - TU Kaiserslautern, Germany}
\address[johannes]{Institut für Informations-, Telekommunikations- und Medienrecht (ITM) - WWU Münster, Germany}
\address[maryam]{Algorithm Accountability Lab (AAL) - TU Kaiserslautern, Germany}

\address[maryam2]{Learning, Data-Analytics and Technology Department, University of Twente, Netherlands}

\begin{abstract}
With the increasing use of AI in algorithmic decision making (e.g. based on neural networks), the question arises how bias can be excluded or mitigated. There are some promising approaches, but many of them are based on a "fair" ground truth, others are based on a subjective goal to be reached, which leads to the usual problem of how to define and compute "fairness". The different functioning of algorithmic decision making in contrast to human decision making leads to a shift from a process-oriented to a result-oriented discrimination assessment. We argue that with such a shift society needs to determine which kind of fairness is the right one to choose for which certain scenario. To understand the implications of such a determination we explain the different kinds of fairness concepts that might be applicable for the specific application of hiring decisions, analyze their pros and cons with regard to the respective fairness interpretation and evaluate them from a legal perspective (based on EU law).
\end{abstract}


\end{frontmatter}

% \linenumbers

\section{Introduction}
\label{section:introduction}

Society as a whole has generally chosen to ban various forms of discrimination from everyday life. This decision is reflected in various laws, such as the German Constitution\footnote{In German called \textit{Grundgesetz}: \href{https://www.gesetze-im-internet.de/gg/BJNR000010949.html}{https://www.gesetze-im-internet.de/gg/BJNR000010949.html}}, the German General Equal Treatment Act \footnote{In German called \textit{Allgemeines Gleichbehandlungsgesetz}:\\ \href{https://www.antidiskriminierungsstelle.de/SharedDocs/Downloads/DE/publikationen/AGG/agg_gleichbehandlungsgesetz.pdf?__blob=publicationFile}{https://www.antidiskriminierungsstelle.de/SharedDocs/Downloads/DE/publikationen/AGG/agg\_gleichbehandlungsgesetz.pdf?\\\_\_blob=publicationFile}} on the German law level or the EU Charter of Fundamental Rights (EU CFR)\footnote{\href{https://eur-lex.europa.eu/legal-content/EN/TXT/?uri=celex:12012P/TXT}{https://eur-lex.europa.eu/legal-content/EN/TXT/?uri=celex:12012P/TXT}} and the General Data Protection Regulation (GDPR)\footnote{\href{https://eur-lex.europa.eu/legal-content/EN/TXT/PDF/?uri=CELEX:32016R0679}{https://eur-lex.europa.eu/legal-content/EN/TXT/PDF/?uri=CELEX:32016R0679}} on the European level.
For several years now, the use of AI-based recommendation systems has been gaining ground in many critical fields of application (e.g. armored drones~\cite{altmann2017autonomous}, predictive policing~\cite{shapiro2017reform} and many more~\cite{narula2018everyday}). However, such systems have the problem that they have to be trained with the help of data that may contain a historical bias, which in turn leads to discrimination when the algorithmic decision making system (ADM system) is in use. 

The goal for a society must be to implement/design fair algorithmic decision making~\cite{muller2020impact}. The term of fairness, however, is multifaceted. Philosophers debate the general question of what fairness means for centuries (see, e.g., the theory of proportional fairness by Aristotle~\cite{ameriks2000aristotle}, but also the works of Rawls~\cite{rawls1971theory} and Dworkin~\cite{dworkin1981equality, dworkin1981equality2} in more recent times). To ensure fairness in machine learning models, which are the basis of AI decision making, it must be mathematically defined so that it can be implemented into such a model. For many possible scenarios of AI decision making, a clear definition of what fairness actually means can only be established by considering the application context~\cite{gurouglu2010unfair}.

The great challenge here is that there exists a large number of measures for determining a bias in principle, of which a large subset is mutually exclusive~\cite{zweig2018fairness}. Different aspects of "fairness" must be evaluated to find the optimal fairness measure for a certain scenario\footnote{The broader topic is so frequently discussed that an entire research community has formed around it, called Fairness, Accountability, and Transparency in Machine Learning (FAT ML)}.
In order to avoid too general formulations and to determine potential solutions on a concrete example, this paper focuses on an elaboration of relevant fairness measures for hiring decisions, groups them, considers their technical/mathematical pros and cons and evaluates the legal compliance and implications in context of the application (based on German and EU law). As far as the paper refers to non member state specific legal provisions, the legal analysis can be transferred to the jurisdiction in other countries.

% ordnung in anzahl fairnessmaße, insbesondere für echtswissenschaftler
% nicht nur it law, sondern auch traditionelle rechtsgebiete können betroffen sein und müssen sich darauf einstellen
%es geht nicht mehr nur im it law, sondern wie man im zivilrecht diskriminierung auffasst. hier gibt es noch nicht so viel kontakt wie im it law
% rechtsdiskussion
\section{AI in hiring decisions}
\label{section:aiinhr}

%There is a wide range of possible applications of AI in the HR environment, covering the whole cycle of employee management: From finding suitable people for a job, even with the help of AI-based background checks\footnote{As of today (2020) there are some start-ups that provide the service of performing AI-based background checks (e.g. Good Egg: \href{https://www.goodegg.io/social-media-screening}{https://www.goodegg.io/social-media-screening}), some analyzing social media information, but some even considering criminal records (e.g. Checkr: \href{https://checkr.com/platform/screenings/criminal-records-check}{https://checkr.com/platform/screenings/criminal-records-check}). Apparently these systems frequently make entity recognition errors, though there seems to be no scientific evidence yet~\cite{url3}(websource).}, to evaluating employee performance, for example as a basis for future hiring decisions~\cite{dattner2019legal}, to estimating which employees have the potential to leave a company in the foreseeable future, either to initiate countermeasures or to avoid expensive training from which the company would not benefit for much longer~\cite{holtom2017today}. While these applications are nothing new in principle, AI not only has the potential to achieve better results, but also to make the process fairer and more objective, e.g. by exposing existing biases, thus allowing such problems to be addressed in a more targeted manner, or even by abstracting from human bias at all.

The idea to use algorithmic decision making as a basis for future hiring decisions~\cite{dattner2019legal}, for example based on an evaluation of employee performance, promises potentially better results and a fairer and more objective process. The promise is that machine learning can help, e.g., by exposing existing biases, thus allowing such problems to be addressed in a more targeted manner, or even by abstracting from human bias at all.

While there seem to be obvious possible benefits of using AI in hiring decisions at first sight, the use of AI in decision making poses a substantial risk of bracing or even strengthening existing biases and discriminatory effects in society. The fundamental idea of decision making by machine learning consists of developing a set of decision rules based on existing data sets, the so-called \textit{training data}~\cite{molnar2018guide}; these decision rules are stored as a so-called \textit{statistical model}. Therefore, any statistical model tends to be structurally as biased against certain social groups as the training data itself, or even tends to emphasize these biases because the predictions of machine learning models are oriented towards the majority group, to be most accurate.

Linked to this is the problem of lacking transparency in the decision-making process, which has a profound implications for the legal system as a whole. AI-based blackbox models, especially artificial neural networks, are hardly explainable~\cite{sokol2020explainability}. The models’ outputs are the result of the weighting of thousands of connections between the initial data set’s features. The hidden layers between input and output layer are extremely hard to examine. The process of producing outputs from the input data is opaque as such. Technologies as, e.g., Shapley Values~\cite{shapley1953value, sundararajan2019many}, LIME~\cite{ribeiro2016should}, Anchors~\cite{ribeiro2018anchors} or Surrogate models~\cite{craven1996extracting} seem promising to lighten up the blackbox. Still, the immense complexity of the billions of connections between the model’s hidden layers cannot be sufficiently explained to this date. This challenges the common process-oriented way of determining discrimination that relies on the actual decision making process, but not its result.

As an example for this processed-oriented assessment, looking at the term “equal treatment” in directive\footnote{Directives together with regulations form the fundamental part of secondary EU law. Unlike regulations, directives are not directly applicable but their provisions must first be implemented by the member states according to their own legal systems.} 2006/54/EC that deals with the implementation of the principle of equal opportunities and equal treatment of men and women in hiring decisions matters\footnote{Directive 2006/54/EC on the implementation of the principle of equal opportunities and equal treatment of men and women in matters of employment and occupation, L 204/23, 26/7/2006 (in the following referenced to as: Equal Treatment Directive).}, the process-based assessment is already hidden in the word \textit{treatment}. The focus of anti discrimination legislation does not lie on the result but the \textit{treatment} of an individual in a specific case, meaning the individual process that caused a result wanted or unwanted by the law. The directive defines ‘direct discrimination’ as a less favorable treatment of one person in comparison to another person on grounds of sex in a comparable actual, former or hypothetical situation.\footnote{Art. 2 (1) lit. a Equal Treatment Directive.} In contrast to that ‘indirect discrimination’ occurs where an apparently neutral provision, criterion or practice would put persons at a particular disadvantage compared to other persons on grounds of sex unless that provision, criterion or practice is objectively justified by a legitimate aim, and the means of achieving that aim are appropriate and necessary.\footnote{Art. 2 (1) lit. b Equal Treatment Directive} The difference between direct and indirect discrimination thus lies in an openly discriminatory \textit{treatment} or process (direct discrimination) vs. a seemingly neutral provision, criterion or practice (indirect discrimination). This means that the same biased result can be based on either a direct or indirect discrimination solely based on the decision process behind the result.

The idea of defining discrimination based on a less favorable treatment of one person in comparison to others can be found in the whole EU anti-discrimination legislation\footnote{Art. 2 Equal Treatment Directive; Art. 2 Directive 2000/43/EC of 29 June 2000 implementing the principle of equal treatment between persons irrespective of racial or ethnic origin, L 180/22, 19/07/2000; Art. 2 Directive 2000/78/EC of 27 November 2000 establishing a general framework for equal treatment in employment and occupation, L 303/16, 02/12/2000; Art. 2 Directive 2004/113/EC of 13 December 2004 implementing the principle of equal treatment between men and women in the access to and supply of goods and services; L 373/37, 21/12/2004.} and is therefore the basis of every member state's anti-discriminatory law. 

The process-based assessment becomes even more evident when looking at the definition of indirect discrimination in detail.\footnote{Compare Art. 2 (1) lit. b Equal Treatment Directive.} Considering for example a job that requires a minimum of physical strength, it is only logical that statistically more men than women are employed due to biological conditions. In case of a lawsuit a judge would need to evaluate whether the formulated requirements are appropriate and the resulting bias is acceptable in this specific case.

This way of thinking is suitable for decision makers who follow written down practices and criteria to be able to make an ideally unbiased decision in an individual case. It is the classic way of applying an abstract rule to an individual case.

As good as this way of thinking may be for humans, as bad or at least not suitable it is for machine learning models. The described process-based way of thinking requires that the actual process is transparent or can be at least rebuilt. In this process-based view, to understand whether a neutral provision resulted in indirect discrimination, it must be made transparent. 
\section{Applicable fairness measures}
\label{section:measures}
The use of a blackbox model, such as an artificial neural network, especially when deep learning is used, excludes the required transparency.
This is one of the reasons why AI is not allowed to make fully automated hiring decisions on its own pursuant to Art. 22 sec. 1 Nr. 1 GDPR. The individual shall not be exposed to the mercy of a technical and opaque process without being able to understand the underlying assumptions and assessment criteria and, if necessary, shall be able to assert their rights and interests.\footnote{See Philip Scholz, in~\cite{datenschutzrecht2019} to Art. 22 GDPR, recital 3.}  

If the decision process cannot be made transparent and explainable, either the idea of a process-oriented assessment collapses or, the new technology cannot be used in those decision making processes where an insight into the decision logic is crucial. %It is no longer possible to say whether a provision is neutral or openly discriminatory and whether it carries an inherent justification, as the provision itself is not transparent but only the result.

In general many known problems revolving around algorithmic discrimination can roughly be divided into two categories: 1) Problems related to the design and deployment of algorithmic decision models and 2) problems related to the training data~\cite{bernstein2018}. While both kinds of problems can make a major contribution to bias in machine learning the first is a matter of established development processes and needs to be refined and improved through experiments and experience while the latter can be mathematically addressed and therefore optimized. What both kind of problems do not address, however, is the question of how to actually define bias. If algorithm-based decision making shall be widely put into practice it is necessary to shift the focus to the definition and assessment of algorithmic bias~\cite{guggenberger2019}. What distribution of results can be deemed discriminatory? It is in trying to answer this question that the idea of using fairness measures is brought up. Since there are over 20 fairness measures~\cite{verma2018fairness} and the discussion about which fairness measures should be used under which circumstances is still in its infancy~\cite{hutchinson201950, friedler2019comparative}, the logical follow-up question must then be, if and how different fairness measures can be integrated into existing legislation.

The technical component of most algorithmic decision making systems is a classification model. Assume that 40 people apply for a vacant position, 10 women and 30 men. A classification model is to decide which of the 40 people are invited to an interview and which are not. The "input" of this model, i.e. the basis on which decisions are to be made, is the collection of characteristics of each applicant that have been converted into numbers (this is referred to as operationalization). The "output" of the model is a recommendation whether a person corresponding to the input should be invited or not. The classification is based on a statistical analysis of historical data, this historical data is called "ground truth". Since there are only two classes "will be invited to the interview" and "will not be invited to the interview", this is a binary classification.

In the context of binary classification, fairness measures now consider one class to be the desired outcome for an individual (called the positive class, e.g. being considered for an interview) and the other class to be the undesired outcome for an individual (called the negative class, e.g. being declined without getting an interview).

\begin{figure}[ht]
    \centering
    \includegraphics[scale=0.7]{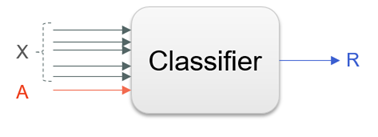}
    \caption{Schematic overview of a classifier.}
    \label{fig:classifier}
\end{figure}

Consider a classification problem for which $\{A_1$, \ldots, $A_m\}$ are sensitive attributes (e.g. gender or religious belief), against which discrimination is unlawful according to anti-discrimination legislation, and $\{X_1$, ..., $X_n\}$ are the remaining attributes. In addition, let $Y$ be the target output (depending on the specific problem either the actual, real-world output or the desired output) and $R$ the system output.
To simplify the problem, we decided to focus only on a binary classification $R = \{0;1\}$ (e.g. gets a hiring recommendation or not) where each feature is only one bit (e.g., is male or not). Also without loss of generality, we assume that there is only one sensitive attribute $A$ which only has two possible values $a_1$ and $a_2$. A schematic overview of such a classifier is shown in figure \ref{fig:classifier}.

An intuitive definition of fair algorithmic decision-making is fairness by unawareness. In this definition, a system is considered fair if it does not use sensitive attributes (i.e. sensitive attributes are not entered into the system for prediction).
It is important to know that fairness cannot be guaranteed by not using sensitive attributes, since in many real-world applications other characteristics of data are correlated with the sensitive attributes~\cite{barocas2018fairness}. Therefore it is absolutely necessary that the information about such attributes is available in order to be able to assess fairness, regardless of whether the values are (may be) relevant in the decision making process by the model or not.

Since there are too many fairness measures to discuss all of them, we limit the perspective to categories of fairness measures, within which the discussion would be very similar due to their matching notion of fairness and compare few example fairness measures for each group with each other.

\begin{enumerate}
    \item Group fairness not based on a ground truth: \\
    In this setting, we assume that we know which of the job candidates have which protected feature, but we do not know the \enquote{ground truth}, i.e., who would be among the best candidates. 
    Under this condition, group fairness measures whether groups were treated equally. It cannot take into account whether the groups were equal from the start, but only whether they were treated the same. So, these measures could be applied to gender, and then evaluate whether both groups as such were treated equally. However, if one of the groups were inherently better than the other, it would disregard this information. Fairness measures not based on a ground truth simply rely on a "fair" distribution of predictions. Therefore, in some way, they must propagate an (e.g., political) goal, as they are always posing quotas that are not reflected in the data set itself. In the scope of this paper we focus on \textbf{Independence} (also known as \textit{statistical parity}, and \textit{demographic parity}) and \textbf{Conditional independence} (also known as \textit{conditional demographic parity} and \textit{conditional statistical parity}~\cite{corbett2017algorithmic}) for this group of fairness measures~\cite{hutchinson201950, barocas2018fairness}. Even though the terms partially match in their expression both measures can lead to hugely different result distributions that justify a separate discussion (see chapter \ref{subsec:independence} and \ref{subsec:conditional_independence}).
    
     \item Group fairness based on a ground truth: \\ Some fairness measures base their computation on a so-called ground truth, which is constructed e.g. by collecting historical data and considered as a reflection of the real world. This inevitably means that historical bias and stereotypes that already exist in today's society influence the decision without reflecting on modern calls for more equality for protected groups. For this category we focus on the fairness measures \textbf{Separation} (as a general term for \textit{equalized odds}, \textit{equalized opportunity} and \textit{conditional procedure accuracy}) and \textbf{Sufficiency} (also known as \textit{conditional use accuracy}, \textit{predictive parity}, and \textit{calibration}~\cite{berk2018fairness, chouldechova2017fair}\footnote{Chouldechova mentions, that the terms \textit{Separation} and \textit{Calibration} are only equal in case of binary decisions.}), due to their different notion of fairness~\cite{hutchinson201950, barocas2018fairness}. While \textbf{Separation} regards the relation of individuals from both, positive and negative class, in the same group, \textbf{Sufficiency} only considers an equal distribution between individuals from the positive class of different protected groups (see chapter~\ref{subsec:separation} and~\ref{subsec:sufficiency}). The different notions can lead to severely different outcomes.   
    
    \item Individual fairness:\\ In contrast, \textbf{Individual fairness} generally means that every individual is treated based on its actual merits, meaning that the treatment of different groups can greatly differ. The idea of the actual fairness measure lies in forming a distance measure that operationalizes the proximity of single individuals' data points in the data set. Since all \textbf{Individual fairness} measures are alike in all their relevant characteristics for the further examination (in particular, they are based on a distance measure and do not require a ground truth) the analysis can be applied to any individual fairness measure (see chapter \ref{subsec:individual_fairness}). Individual fairness measures and group fairness measures collide with each other, both cannot be accomplished to full extent at the same time.
    
    \item Counterfactual fairness:\\ There are few fairness measures that do not fit in any of those groups. Unlike the fairness measures mentioned above, \textbf{Counterfactual fairness} is based on sociopolitical assumptions instead of operationalizable mathematical functions (see chapter \ref{subsec:counterfactual_fairness}). Some researchers argue that measures which are based purely on probabilistic independence cannot sufficiently consider social biases and are unable to address how unfairness is occurring in the task at hand~\cite{kusner2017counterfactual, kilbertus2017avoiding}.
\end{enumerate}
While for independence, separation, and sufficiency other terms can be found as well, we decided to reference the terms used by ~\cite{barocas2018fairness} as it addresses the problem of multiple names for the same concepts. The names are used in other recent publications as well~\cite{hutchinson201950}. There are also fairness measures that explicitly evaluate a ranking (e.g. \cite{singh2018fairness, beutel2019fairness}, which can also be used in hiring decisions. However, since any ranking can be mapped to a binary decision by setting a threshold, these are not additionally discussed below.

\subsection{Independence}
\label{subsec:independence}
One of the most common mathematical definitions of fairness is the definition of independence. According to \textbf{Independence}, the system output must be independent of sensitive attributes~\cite{dwork2012fairness}. The probability that a data instance is predicted by the system to be in the positive class should be the same for the different groups identified by the sensitive attributes.

$$
Pr\{R=1|A=a_1\}=Pr\{R=1|A=a_2\}
$$

\begin{figure}[h]
    \centering
    \includegraphics[width=0.2\textwidth]{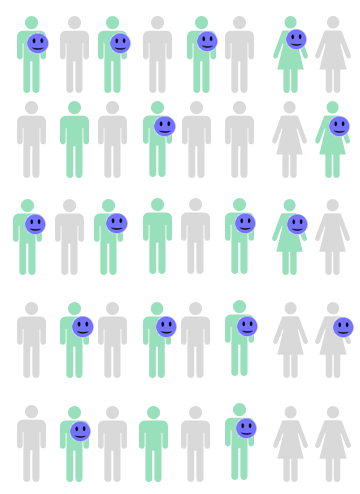}
    \caption{Example of a job hiring system that is fair according to the definition of \textbf{Independence}. Green individuals are actually qualified for the job and people with blue badges are accepted for the job by the system. The probability of acceptance of a person is independent of their gender and is 40\%.}
    \label{fig:independence_example}
\end{figure}

To better understand this definition, consider another example of a job hiring system. Imagine that 40 people applied for the job, including 10 women and 30 men. Among them, 50\% of men and 30\% of women are actually qualified for the job.

$$
Pr\{R=accepted|A=woman\}=Pr\{R=accepted|A=man\}
$$
If the system accepts the same percentage of men as women to be invited for interview, it is considered fair. Figure~\ref{fig:independence_example} shows an illustrating example of a job hiring system in which the condition of \textbf{Independence} is satisfied.

It should be noted that the percentage of qualified persons in the men and women group is not the same, which may result in some qualified men who are not accepted for the job or some women who are not being qualified enough and still get accepted for the job even though the fairness measure is fulfilled. Therefore, this definition of fairness may not be appropriate in certain social situations. In some cases though, it may be a good choice, for example, if one’s goal is to achieve an equal number of men and women at a job. On the contrary, for most application cases the qualification of an applicant will be one of the decisive factors, if not the only. In these cases \textbf{Independence} is no proper definition of fairness.

\subsection{Conditional Independence}
\label{subsec:conditional_independence}

\textbf{Conditional Independence} is another definition of fairness which refers to the equal probability of assigning instances to the positive class for a part of both groups which are similar in some other attributes X. 
$$
  Pr\{R=1|A=a_1,X_i=x_{i},...,X_j=x_{j}\} = Pr\{R=1|A=a_2,X_i=x_{i},...,X_j=x_{j}\}  
$$
For example, for the job hiring system, we can consider a certain threshold of university GPA (Grade Point Average) as the conditional attribute. Then, the system is fair if the ratio of the accepted persons from those applicants who surpass the threshold GPA value is the same for both groups.
Assume 1.000 applicants applied for a job, 400 women and 600 men. From them, 104 persons are accepted by the system, 32 women, 72 men. Assume 60\% of men and 40\% of women surpass the GPA threshold. If 72 from the 360 men who have good GPAs (20\%) are accepted for the job, and 32 from the 160 women who have good grades (20\%) are accepted for the job, the system satisfies:
$$
Pr\{R=Accept|A=man, GPA=good\} = Pr\{R=Accept|A=woman, GPA=good\}
$$

If this condition is also satisfied for all possible values of GPA, the system is fair based on \textbf{Conditional Independence}.  

This means that if the applicants are divided based on the different values of GPA, then a system is fair if the proportion of the accepted people for women and men within one GPA group is the same. In other words, a male and a female with the same GPA have the same chance of being accepted for the job. 

One challenge is that if the conditional attribute is not binary or even worse continuous, the number of people who fall in each group may be very low, so that it may be hard to satisfy this fairness criterion. Generally, it is possible to discretize it by dividing the range into two or more intervals and to check whether the conditional probability equation holds for all discrete values. For example instead of checking GPA values as {4.0, 3.7, 3.3, 3.0, 2.7, 2.3, 2.0, 1.7, 1.3, 1.0} for \textbf{Conditional Independence} only two values like good and bad can be considered.

\subsection{Separation}
\label{subsec:separation}
\textbf{Separation} is satisfied, if the system output (prediction) is independent of the sensitive attribute, provided that the desired output is known~\cite{hardt2016equality, berk2018fairness}. An ADM system is considered fair according to \textbf{Separation} if the following two conditions are met:

\begin{enumerate}
    \item The probability that an instance is assigned to the positive class under the condition that the instance actually belongs to the positive class should be the same for both groups. This implies equal true-positive rates\footnote{See appendix} for the groups. 
$$
Pr\{R=+|Y=+,A=a_1\}=Pr\{R=+|Y=+,A=a_2\}
$$
    \item The probability that an instance is assigned to the positive class under the condition that the instance actually belongs to the negative class should be the same for both groups. This implies equal false-positive rates\footnote{See appendix} for the groups. 
$$
Pr\{R=+|Y=-,A=a_1\}=Pr\{R=+|Y=-,A=a_2\}
$$
\end{enumerate}

Mapped onto an AI based job hiring system this means the first equation is fulfilled if the chance of accepting individuals who are actually qualified for the job is the same for both groups.
$$
Pr\{R=Accept|Y=qualified,A=woman\}=Pr\{R=Accept|Y=qualified,A=man\}
$$

The second equation is fulfilled if the chance of accepting individuals who are actually not qualified for the job is the same for both groups as well.
$$
Pr\{R=Accept|Y=unqualified,A=woman\}=Pr\{R=Accept|Y=unqualified,A=man\}
$$

\begin{figure}[h]
    \centering
    \includegraphics[width=0.2\textwidth]{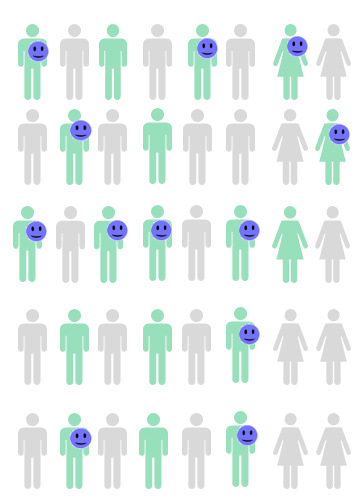}
    \caption{Example of a job hiring system which is fair based on the definition of \textbf{Separation}. Green individuals are actually qualified for the job and people with blue badges are accepted for the job by the system. The probability of acceptance for any person who is really qualified is independent of their gender and is 66\%. Moreover, the probability of acceptance for any person who is actually not qualified is independent of their gender and is 0\%.}
    \label{fig:separation_example}
\end{figure}

Figure \ref{fig:separation_example} shows an illustrating example of a job hiring system in which both conditions of \textbf{Separation} are satisfied.

Note that, although in this example the formulas of fairness are similar for \textbf{Separation} and \textbf{Conditional Independence}, they refer to different notions. The condition part in \textbf{Conditional Independence} is regarding some feature or characteristic of the applicants (e.g., GPA). However, in \textbf{Separation}, the condition is on the desired label (ground truth). Hence, for the job hiring example, to compute \textbf{Separation} one needs to know the correct decision for each person ($Y$), i.e., whether a person is actually qualified (here, by being qualified, we mean that the correct decision for him/her is "to be accepted for the job"). \textbf{Conditional Independence} does not use $Y$ or any desired output and only considers the features of applicants.

\textbf{Separation} promotes both equal bias and equal accuracy in all demographic groups and thus discriminates in models that perform well only on the majority~\cite{hardt2016equality}. Therefore the measure helps to further strengthen existing bias in the data set (the ground truth). 

Considering the explanation of \textit{McNarma et al.} this perspective makes sense: \textit{[...] an algorithm that reflects this difference is not 'unfair' but is rather a reflection of real underlying differences. [...] surely we would not want to label 'unfair' a prediction algorithm which is perfectly accurate! The job of the algorithm is to predict the world as it is; changing the world is out of scope}~\cite{mcnamara2019equalized}. If the results were to be differentiated between groups, there would be a risk of discrimination in situations where the data collection process systematically discriminates against a group~\cite[p.695ff]{barocas2016big}. However, the gap between two groups will tend to be enlarged over time if the measure is based on a ground truth that may be heavily biased or incomplete. The question of whether and when positive discrimination should be used to reduce inequality is one of the central issues in legal debates.

\subsection{Sufficiency}
\label{subsec:sufficiency}
\textbf{Sufficiency} is another fairness criterion, which considers a ground truth. Based on this definition of fairness, an ADM system is fair, if the desired output is conditionally independent of the sensitive attribute, given the system output.

For a binary classifier, this criterion can be formally expressed by the following two conditions:

\begin{enumerate}
    \item The probability that an instance belongs to the positive class on the condition that it is assigned to the positive class should be the same for both groups. This implies equal precision\footnote{See appendix} for the groups. 
\begin{equation*}
Pr\{Y=+|R=+,A=a_1\}=Pr\{Y=+|R=+,A=a_2\}
\end{equation*}

    \item The probability that an instance belongs to the positive class on the condition that it is assigned to the negative class should be the same for both groups. This implies an equal false omission rate\footnote{See appendix} (FOR) for the groups. 
\begin{equation*}
Pr\{Y=+|R=-,A=a_1\}=Pr\{Y=+|R=-,A=a_2\}
\end{equation*}
\end{enumerate}

Mapped onto an AI based job hiring system this means the first equation is fulfilled if the percentage of actually qualified individuals for the job among those accepted by the system is the same for both groups.
$$
Pr\{Y=qualified|R=accepted,A=woman\}=Pr\{Y=qualified|R=accepted,A=man\}
$$

The second equation is fulfilled if the percentage of actually qualified individuals for the job among those not accepted by the system is the same for both groups.

$$
Pr\{Y=qualified|R=rejected,A=woman\}=Pr\{Y=qualified|R=rejected,A=man\}
$$

\begin{figure}[h]
    \centering
    \includegraphics[width=0.2\textwidth]{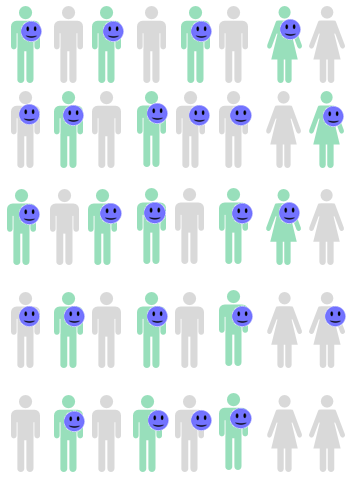}
    \caption{Example of a job hiring system which is fair based on the definition of \textbf{Sufficiency}. Green individuals are actually qualified for the job and people with blue badges are accepted for the job by the system. The probability of being actually qualified for any individual who is accepted for the job is the same for both groups and it is equal to 75\%. Moreover, the probability of being actually qualified for any individual who is rejected for the job is the same for both groups and it is equal to 0\%.}
    \label{fig:sufficiency_example}
\end{figure}

Figure~\ref{fig:sufficiency_example} shows an illustrating example of a job hiring system in which both conditions of \textbf{Sufficiency} are satisfied.

At first glance, the idea behind \textbf{Sufficiency} to move towards equal error rates embodies the fundamental principle of equal treatment. However, the mathematical formula behind \textbf{Sufficiency} allows the false-positive rates to diverge between groups, which can have a severe impact and makes this fairness measure highly debatable.

Except in degenerated cases, the fairness measures \textbf{Independence}, \textbf{Separation} and \textbf{Sufficiency} are mutually exclusive \cite{barocas2018fairness}.

\subsection{Individual fairness}
\label{subsec:individual_fairness}
Based on \textbf{Individual fairness}, a system is considered fair if the outputs of the system are similar for similar individuals~\cite{dwork2012fairness}. This definition needs to consider a distance measure between individuals and a distance measure for identifying the similarity between system outputs. In other words, an ADM system is considered fair according to \textbf{Individual fairness} if the following condition is met:

$$
D(O(I_i),O(I_j))<d(I_i,I_j)
$$

\begin{figure}[h]
    \centering
    \includegraphics[scale=0.7]{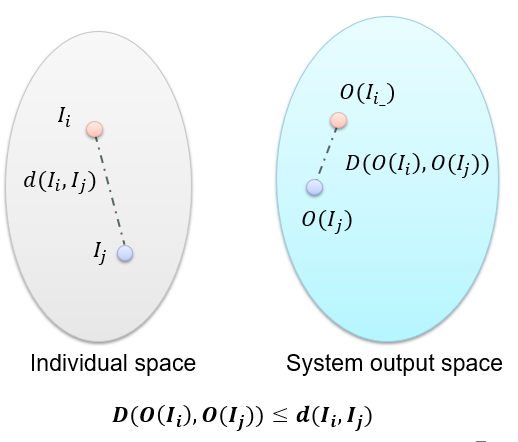}
    \caption{Individual fairness.}
    \label{fig:individual_fairness}
\end{figure}

Where $d$ is the distance measure between two individuals and $D$ is the distance measure between the outputs $O$ of the system for any two individuals $I$ (see Figure~\ref{fig:individual_fairness}.

\subsection{Counterfactual fairness}
\label{subsec:counterfactual_fairness}

A system satisfies the counterfactual fairness criterion, if the system output $R$ (its probability distribution) for any member of a group (i.e., race, gender, sexual orientation), is the same as when he or she is from a different group. To check this criterion causal relationships between attributes, expressed by the structural equations, can be depicted by a causal graph. By assuming such a causal graph between the attributes, a system is counterfactually fair if the following condition is satisfied~\cite{kusner2017counterfactual}:

%https://arxiv.org/pdf/1703.06856.pdf
%https://link.springer.com/content/pdf/10.1007/s11023-020-09529-4.pdf

\begin{equation*}\label{eq:counterfact}
\Pr(R_{A\leftarrow a_1}=r| X=x, A=a_2)=\Pr(R_{A\leftarrow a_2}=r|X=x, A=a_2) 
\end{equation*}

where $R_{A \leftarrow a_2}$ is the actual prediction and $R _{A\leftarrow a_1}$ is the counterfactual prediction, by the imagination that $A = a_1$ (while actually $A = a_2$ is true). $X$ represents the other variables in the causal graph.

In the causal graph the nodes denote the attributes and edges represent the causal relationships between them. Mapped onto the job hiring system, the chance that a woman with the profile $\{X_1=x_1,\dots,X_n=x_n\}$ gets accepted for the job is $Pr\{R_{A \leftarrow woman}=Accept|A=woman, X_1=x_1,\dots, X_n=x_n\}$. If attribute $A$ is changed to $man$ in the causal graph, the value of some attributes might change as well according to their causal relations in the graph, while some others remain fixed. The chance that this counterfactual pair is accepted for the job is $Pr\{R_{A \leftarrow man}=Accept|A=woman, X_1=x_1,\dots, X_n=x_n\}$. When both probabilities are the same the system is considered counterfactually fair. 

\begin{figure}[h]
    \centering
    \includegraphics[width=0.4\textwidth]{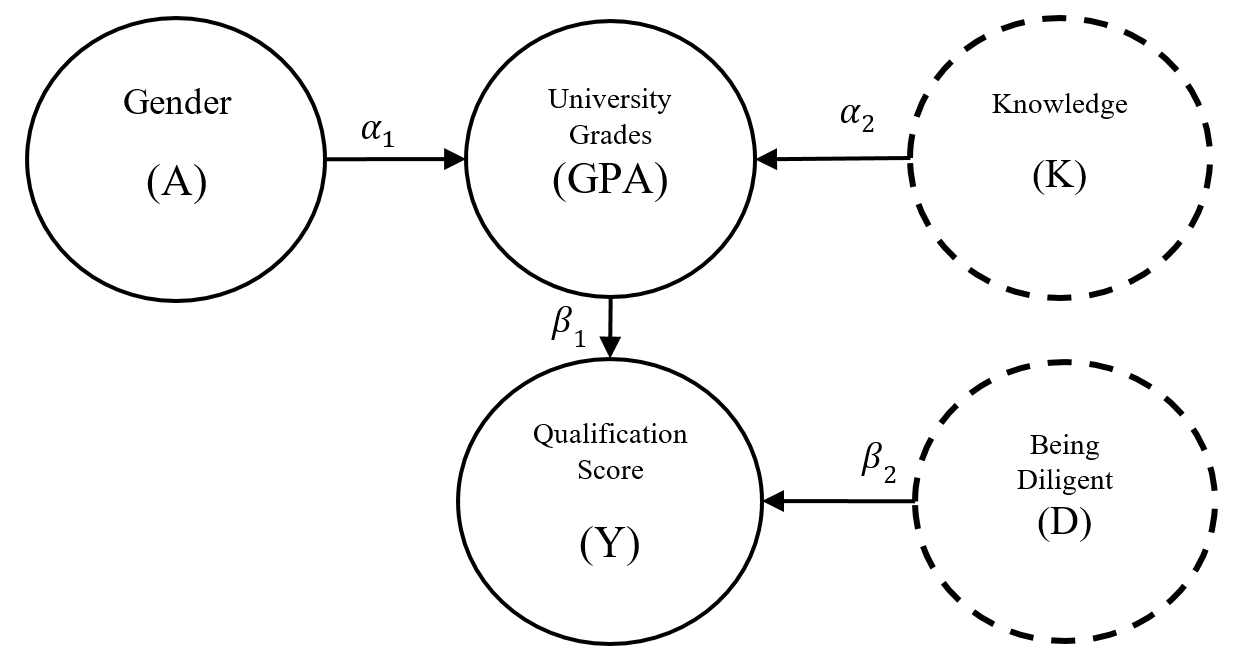}
    \caption{Example of a simple causal graph for a job hiring system. }
    \label{fig:counterfactual}
\end{figure}

Counterfactual fairness can be further explained by the following illustrative example. Figure~\ref{fig:counterfactual} depicts a simplified example of a causal graph for a job hiring system. The example has been inspired by the example of~\cite{kusner2017counterfactual} and adapted according to an example from~\cite{makhlouf2020applicability}. The directed edges show the causal relationships between the attributes. Here, two observable variables are considered, the gender and the average of university grades. Moreover, in this graph, two hidden or latent variables are considered, the \textit{Knowledge} (K) and \textit{Being Diligent} (D). These variables are not dependent on any observable attributes. 

Based on this causal graph, the output \textit{qualification score} (Y) depends on the observable attribute \textit{average of the university grades (GPA)}, and the hidden or latent attribute \textit{being diligent} ($D$) indicating how much an applicant is diligent. Moreover, the attribute \textit{GPA} depends on the observable sensitive attribute $A$ which depicts the gender and a hidden attribute \textit{knowledge} $K$ that represents how much knowledge the applicant has. This causal graph can be specified by two equations as follows. 

\begin{align*}
    GPA=\alpha_1 \cdot A+\alpha_2 \cdot K\\
    Y= \beta_1 \cdot GPA+\beta_2 \cdot D
\end{align*}

In practice the coefficients can be learned from the data by defining an arbitrary joint probability distribution for all latent variables. A specific distribution for the hidden variables is assumed, based on which the coefficients of the causal model are learned in a way that leads to the best possible fitting to the data for example by the maximum likelihood method, though there are some additional approaches. To find a fitting causal model enough data from all sensitive groups is required. For further explanation the coefficients are set to $\alpha_1=0.2$, $\alpha_2=0.8$, $\beta_1=0.4$ and $\beta_2=0.6$.  

Considering the causal model, the male applicant Bob  ($A_\text{Bob} = 1$), has a normalized GPA of $0.75$ ($GPA^{Bob}=0.75$), where $GPA=\{x \mid x\in[0,1], x\in\mathbb{Q}\}$. The predicted qualification score $Y$ for Bob given by the system is equal to 0.7 ($Y^{Bob}=0.7$). 

To check for \textbf{Counterfactual Fairness} now Bob’s qualification score has to be computed under the assumption that his gender is female ($A^{Bob}=0$) including all causal consequences on the other attributes as depicted by the causal graph.  
At first, the corresponding value of the latent attribute $K$ should be computed using the known information ($A^\text{Bob}= 1$,  $GPA^\text{Bob}= 0.75$, $Y^{\text{Bob}}_{A\leftarrow 0} = 0.7$). To find $K^\text{Bob}$ and $D^\text{Bob}$ the structural equations can be used as follows:

\begin{align*}
    K^\text{Bob}=\frac{GPA^\text{Bob}-\alpha_1 \cdot A^\text{Bob}}{\alpha_2}=\frac{0.75-0.2 \cdot 1}{0.8}=0.69\\
    D^\text{Bob}=\frac{Y^\text{Bob}-\beta_1 \cdot GPA^\text{Bob}}{\beta_2}=\frac{0.7-0.4 \cdot 0.75}{0.6}=0.67 
\end{align*}

For comparison with the counterfactual entity, the sensitive attribute $A$ is set to zero, assuming that Bob is female. The other variables are updated according to the structural equations.

\begin{align*}
    GPA^\text{Bob}_{A\leftarrow0}=\alpha_1\cdot0+\alpha_2\cdot K^\text{Bob}=0.2\cdot0+0.8\cdot0.69=0.55
\end{align*}

Now the output $R^\text{Bob}_{A\leftarrow 0}$ of the actual model is computed for the counterfactual pair as well. To assess \textbf{Counterfactual Fairness} of a model, the predicted outputs need to be the same in the actual and counterfactual worlds for every possible individual. Since this criterion is extremely hard to fulfill, Russel et al. propose to compare the distribution of the actual and the counterfactual data by applying a loss function which must not exceed a certain threshold (therefore called \textit{approximate counterfactual fairness}~\cite{russell2017worlds}) to be specified e.g. by a regulating instance. 

The main challenge of \textbf{Counterfactual Fairness} in
real-world systems is the construction of the causal graph. A common procedure is to look at several causal graphs, then fit them to the data (calculate the coefficients) and select the one that seems to be best fitted~\cite{kusner2017counterfactual}. When creating causal models, strong assumptions are usually made, especially in the context of counterfactual claims~\cite{dawid2000causal}. Moreover, counterfactual assumptions such as structural equations are generally not falsifiable, even when there are additional test data for comparison. This is because there are many structural equations that are compatible with the same observable distribution~\cite{pearl2000causality}. Therefore, these causal models should be developed by domain experts to the best of their knowledge. In addition, such models should be considered to be temporal and should be updated repeatedly if new data contains information that contradict the currently adopted model~\cite{kusner2017counterfactual}.

\section{Legal perspective}
\label{section:legal}

There is no clear answer to what equality actually means or when it is accomplished. The term "fairness" seems to be connected to the notion of equality in demanding that people of all different kinds are either treated equally or distributed equally in a decision output set or some mix of both. The term fairness implies both, equality and freedom rights, for individuals as well as for groups. It connects the different social layers of individuality and personality on the one hand and group behavior and treatment on the other. To date, there has been no legal definition of fairness beyond individual decisions of jurisdiction~\cite[p.5]{wachter2020fairness}. This is mainly due to the process-oriented assessment of equality. With the upcoming algorithm-based decision making though, it will be essential to find a clear specification of fairness as the assessment will no longer work on a process level. Rather, there will be a decisive shift to a result-based assessment~\cite{guggenberger2019}. It is therefore of utmost importance to analyze, whether and how the fairness measures presented in this paper comply with EU anti-discrimination legislation.

\subsection{Relevant Legislation}
\label{subsec:relevant_legislation}
Before the integration of fairness measures into existing anti-discrimination legislation is examined, the relevant legislation for said fairness measures in algorithm based decisions must be specified.

This poses a problem as many different legal acts influence each other in this domain. On a German constitutional level there is Art. 3 section 3 Basic Law \footnote{In Germany called \textit{Grundgesetz}} (BL) that especially prohibits any discrimination based on sex, parentage, race, language, homeland and origin, faith, religious or political opinions and disability. In principle it is the Basic Law that is standing on top of all other German laws and that all other laws must comply with. However, according to Art. 23 BL in combination with Art. 4 sec. 3 Treaty on the European Union (TEU) and based on a judgment of the European Court of Justice (ECJ), all legislation of the European Union, primary as well as secondary law, enjoys primacy of application, even to constitutional law.\footnote{CASE C-6/64 \textit{Costa v E.N.E.L.} EU:C:1964:66 [1964] ECR- I00585; see instead of many Zuleeg, Das Recht der Europäischen Gemeinschaften im innerstaatlichen Bereich, pp. 136; Streinz, Europarecht, Notice 220; Beljin, EuR 2002, 351 (353).} As such, European Law prevails over Art. 3 BL.

Crucial for the domain of anti-discrimination are therefore Art. 21 of the Charter of Fundamental Rights of the European Union (EU CFR), Art. 19 and 157 sec. 4 TFEU and especially the four Directives 2000/43/EC\footnote{Directive 2000/43/EC of 29 June 2000 implementing the principle of equal treatment between persons irrespective of racial or ethnic origin, L 180/22, 19/07/2000.}, 2000/78/EC\footnote{Directive 2000/78/EC of 27 November 2000 establishing a general framework for equal treatment in employment and occupation, L 303/16, 02/12/2000.}, 2004/113/EC\footnote{Directive 2004/113/EC of 13 December 2004 implementing the principle of equal treatment between men and women in the access to and supply of goods and services; L 373/37, 21/12/2004.} and 2006/54/EC\footnote{Directive 2006/54/EC on the implementation of the principle of equal opportunities and equal treatment of men and women in matters of employment and occupation, L 204/23, 26/7/2006.}. The German lawmaker has implemented the directives by enacting the General Equal Treatment Act (GET)\footnote{In Germany called \textit{Allgemeines Gleichbehandlungsgesetz}, 2016-08-14 (Federal Law Gazette I, page 1897)}. It prohibits discrimination between private subjects based on the protected features of race or ethnic origin, gender, religion or belief, disability, age or sexual orientation.

The General Equal Treatment Act as the German implementation of EU anti-discrimination law thus represents the most important legislative act in anti-discrimination legislation in Germany, especially due to its direct effect on private subjects. Unlike the different directives named above, the GET constitutes a comprehensive anti-discrimination law that applies to a broad field of cases. As it is further a direct implementation of EU law and its provisions, it must be construed in light of and in conformity with EU primary and secondary law. This paper will thus treat the question of the integration of fairness measures into anti-discrimination law on the basis of the German General Equal Treatment Act. In doing so, the act must, as stated, at all times be construed in the light of EU primary and secondary law and is therefore representative of the directives' and EU CFR, TFEU and TEU's content.

\subsection{What Fairness Measures are applicable to the GET?}
\label{subsec:apllicability}
The GET explicitly lists HR scenarios of hiring or promotion inside its scope pursuant to Section 2 (1) No. 1 and 2 GET. According to Section 7 (1) GET, employees shall not be permitted to suffer discrimination on any of the grounds named above. Section 6 (1) sentence 2 GET states that job applicants shall also be considered “employees”.
The decisive question thus is what is understood as discrimination under the GET. Section 3 GET contains several definitions of discrimination. However, the definitions in Section 3 (3-5) GET referring to “harassment” require a conduct\footnote{See Alex Baumgärtner, in~\cite{beckogk2020} to section 3 GET, recital 110.} and thus a human behavior. They are therefore irrelevant for algorithmic decision making.

Instead, the focus for algorithmic decision making lies on the first two paragraphs that define direct and indirect discrimination. Direct discrimination is presumed when one person is treated less favorably than another is, has been or would be in a comparable situation on any of the grounds named above. In contrast, indirect discrimination is presumed when an apparently neutral provision, criterion or practice would put persons at a particular disadvantage compared with other persons on any of the grounds named above, unless that provision, criterion or practice is objectively justified by a legitimate aim and the means of achieving that aim are appropriate and necessary.

Unlike the so called direct discrimination in paragraph 1, indirect discrimination can be inherently justified, so that it can no longer be regarded as discrimination in the sense of the GET. 

As shown before, the difference between direct and indirect discrimination relies on a process-oriented assessment that looks at the actual decision process and can no longer be upheld for algorithmic decision making. It is therefore necessary to find a unified definition of discrimination that does not differentiate between direct and indirect discrimination. This unified definition must be based on the common properties of both direct and indirect discrimination. The common ground between the two forms of discrimination is that in both cases a person is treated less favorably than another is, has been or would be treated in a comparable situation on any of the grounds stated in Section 1 GET.

How such a unified definition complies with the GET and the underlying directives remains unclear to this date. It is unlikely that such a definition can be simply derived from the GET’s or the underlying directives’ text. Even when factoring in the primary EU law of EU CFR and TFEU, it is doubtful that they allow for such a wide interpretation. It will therefore indeed be necessary to establish a new discrimination term in the EU directives to adapt to the new situation of algorithmic decision making.

Since the development of such a term remains at least partly a political decision~\cite{wong2019democratizing}, it exceeds the scope of this paper. Nevertheless, as the new term is intended to unify the common features of both direct and indirect discrimination, different fairness measures and their respective characteristics can be considered in the context of the term of \textit{less favorable treatment} from a legal perspective.

The GET follows the idea of defining discrimination by comparing a potentially discriminated person with a legitimate comparator group. The comparator group is determined based on the reach or scope of the respective algorithmic decision. For example, there is no sense in comparing the treatment of a person in Germany with the treatment of persons in China, when the reach of the algorithmic decision apparently is Germany-wide and not global. A thorough discussion about the problems of defining a legitimate comparator group for global reach has already been performed by \textit{Wachter et al.}~\cite[p.12]{wachter2020fairness}.

Having determined a legitimate comparator group, an actual, former or hypothetical less favorable treatment of the potentially discriminated person compared to the comparator group must be identified. It is this requirement where fairness measures can be integrated into the assessment. The GET and the respective directives that it implements do not state what a less favorable treatment encompasses. This is surprising as a statistical definition of unequal treatment is inevitable to establish indirect discrimination where a direct causality between conduct and discrimination cannot be proven. It is made even more evident when one regards the fact that indirect discrimination is built upon the idea of \textit{disparate impact} in US anti discrimination law that directly looks at a statistical distribution\footnote{US Supreme Court, Griggs v. Duke Power Co., 401 U.S. 424 (1971), p. 431 and  "Uniform guidelines on employee selection procedures": \href{http://www.uniformguidelines.com/uniform-guidelines.html\#18}{http://www.uniformguidelines.com/uniform-guidelines.html\#18}(last accessed 28/08/2020)}~\cite[p.9]{zuiderveen2019algorithmic}~\cite[pp.178]{fba2011discrimination}. However, judges have rather preferred to reason directly about influences using a mixture of common sense and direct reasoning about the impact of contested rules~\cite[p.49]{wachter2020fairness}. In addition, various fairness measures have been applied inconsistently across member states to assess indirect discrimination~\cite[p.37]{wachter2020fairness}.

Still, the ECJ has seen the need to find a statistical measure for indirect discrimination and in consequence stated that indirect discrimination (between men and women) is given when a seemingly neutral rule puts considerably more workers of one sex at a disadvantage than the other.\footnote{CASE C-363/12, \textit{Z. v A Government department and The Board of management of a community school}, EU:C:2014:159 [2014] ECR- I159, paragraph 53; CASE C-1/95 \textit{Hellen Gerster v Freistaat Bayern} EU:C:1997:452 [1997] ECR- I05253, paragraph 30, 35; CASE C-123/10 \textit{Waltraud Brachner v Pensionsversicherungsanstalt} EU:C:2011:675 [2011] ECR- I10003, paragraph 56; CASE C-7/12 \textit{Nadežda Riežniece v Zemkopības ministrija and Lauku atbalsta dienests} EU:C:2013:410 [2013], paragraph 39; CASE C-527/13 \textit{Lourdes Cachaldora Fernández v Instituto Nacional de la Seguridad Social (INSS) and Tesorería General de la Seguridad Social (TGSS)} EU:C:2015:215[2015], paragraph 28; CASE C-385/11 \textit{Isabel Elbal Moreno v Instituto Nacional de la Seguridad Social (INSS) and Tesorería General de la Seguridad Social (TGSS)} EU:C:2012:746 [2012], paragraph 29; CASE C-256/01 \textit{Debra Allonby v Accrington \& Rossendale College, Education Lecturing Services, trading as Protocol Professional and Secretary of State for Education and Employment.} EU:C:2004:18 [2004] ECR- I00873, paragraph 79; CASE C‑83/14 \textit{"CHEZ RazpredelenieBulgaria" AD v Komisia za zashtita ot diskriminatsia} EU:C:2015:480 [2015], paragraph 101; CASE C-161/18 \textit{Violeta Villar Láiz v Instituto Nacional de la Seguridad Social (INSS) and Tesorería General dela Seguridad Social (TGSS)} EU:C:2019:382 [2019] paragraph 38 seq.; CASE C-300/06 \textit{Ursula Voß v Land Berlin} EU:C:2007:757 [2007] ECR-I10573, paragraph 42.} This measure is called \textit{negative dominance} by \textit{Wachter et al.} and requires a two step assessment: (1) the majority of the disadvantaged group, meaning the group of people for which the respective norm or process has a negative consequence, belongs to a protected class (in the sense of Section 1 GET), and (2) at the same time only a minority of the protected class is part of the advantaged group \cite[p.50]{wachter2020fairness}.

Yet, this statistical measure is in no way further legitimized by the ECJ. The ECJ has simply adopted this specific fairness measure without considering other fairness measures, especially those based on a ground truth such as for example \textbf{Separation}. Still, it is not clear whether the standard adopted by the ECJ is the best or the only standard applicable in, or compliant with, anti-discrimination law. As shown, several fairness measures are mutually exclusive which gives even more reason to analyze the possible integration and implications of other fairness measures than the ECJ’s into the GET and into anti-discrimination law. 

The analysis of fairness measures shows that fairness is always linked to comparison. EU anti-discrimination law allows the potentially discriminated person to - in addition to referring to a current treatment - prove a discrimination by claiming a less favorable treatment in comparison with a former or hypothetical treatment. This means that an employer who has to decide on hiring two identical people for a job and decided to hire one of them in the past, but did not hire the other one now has to justify this decision in front of the court.

For algorithmic decision making, however, this comparison with historical decisions is only possible to a limited extent. A comparison between individuals can only be performed by comparing the differences of their relations to the respective data set they are part of~\cite{guggenberger2019}. Machine learning models are trained with training data sets with which the model learns its prediction. The trained model is then making predictions for new data without being trained anymore. As such, the decision process is defined by the structure of the training set and the conditions set for the model during training. This means that a discrimination of a person cannot be based on a differing individual opinion or preference in the past, because the machine learning model always decides based on a determined algorithm that has been developed from a certain training data set.\footnote{The typical idea in anti-discrimination law of altering individual preferences can only be established by training a model with new data which leads to a change of their decision processes.} Therefore, the idea of comparing hypothetical and former treatments in anti discrimination law has to be partly reconceived.

In case of algorithmic decision making, investigating accusations of unequal treatment means investigating statistical values that depend on the respective data set. Even when considering \textbf{Individual Fairness} the definition of what can be considered a high similarity depends on the similarity to other entities in the data set.

The strong dependence on statistical analysis shows the necessity of a legal discussion of possible fairness measures for algorithmic decision making. These fairness measures define whether an alleged discrimination is given or not. A discussion which fairness measure can be legally compliant in which scenario is therefore inevitable.

\subsection{Compliance of fairness measures}
\label{subsec:compliance}
Which fairness measures are able to inversely define  the term \textit{less favorable treatment} in conformity with the GET remains unclear. Several fairness measures could be applicable. There is however, a decisive constraint: the fairness measures must comply with the GET, the underlying directives and foremost with EU primary law, especially the EU CFR. 

The following analysis of the different fairness measures will only look at a scenario of a binary prediction with a binary protected feature for simplification.

\subsubsection{Independence}
\label{chap:independence}
\textbf{Independence} is maybe the most drastic of all fairness measures. Not based on ground truth it simply demands the same distribution of prediction between the protected groups. As long as this quota is fulfilled by the algorithm, the result distribution is deemed fair. The fairness measure does not consider the ground truth and by that, the statistically captured qualification of the single individuals in the data set. This means that a fix distribution goal is prioritized over the individual qualification of people. If the most qualified applicants from one group and random people from the other group are hired, \textbf{Independence} can still be achieved. The fixed quota serves as a means of affirmative action\footnote{Affirmative actions mean policies or provisions that support groups of people that have been usually discriminated against in society. They often infer a preferential selection of people from protected groups to increase their share in the overall decision distribution. A typical example are quotas for women in supervisory boards.} that tries to work against structural inequalities in the society, such as generally lower level of education for people with darker skin~\cite{wexler2020knowledge}. The idea of affirmative action is well known to anti-discrimination law. Above all Art. 23 EU CFR allows for \enquote{measures providing for specific advantages in favour of the under-represented sex}. Based on that, Art. 157 sec. 4 TFEU more specifically allows for \enquote{measures providing for specific advantages in order to make it easier for the underrepresented sex to pursue a vocational activity or to prevent or compensate for disadvantages in professional careers}. In more general terms, section 5 GET furthermore allows for positive discrimination to work against existing disadvantages due to a protected feature of a person. However, following a quota inside the discrimination definition would mean a lot more than \enquote{specific measures} or a subsequent justification of said positive discrimination. By making a quota an inherent part of the discrimination definition, a situation of actual positive discrimination and affirmative action would no longer be considered discrimination as the fairness measure of \textbf{Independence} would not allow to establish a \enquote{less favorable treatment} if the quota is met. Instead of presenting specific measures in certain decision areas, affirmative action would become the norm, not the exception.

It is highly doubtful if such negligence of individual qualification is still constitutional or compliant with primary EU law. Art. 157 sec. 4 TFEU and Art. 23 EU CFR show that measures of affirmative action are considered an exemption that requires a special competence and are not supposed to be widely implemented such as in a general discrimination term. Positive discrimination shall only be a punctual measure. By stating that the principle of equality shall not prevent adoption of affirmative measures, both Art. 23 EU CFR and Art. 157 sec. 4 TFEU imply that such measures do not establish a standard of equality themselves but rather are an exception that the law explicitly allows for. 

The reason why positive discrimination cannot be seen as an equality standard lies in the liberal idea of equality of opportunity. Fixed quotas, which are not based on ground truth are not suitable for giving everyone the same chances of success, but rather ascertain the status of still being within a specific percentage that must be met~\cite[p.7]{dwork2012fairness, hardt2016equality}. EU law follows the aforementioned liberal idea of equality of opportunity. The European primary law can be seen as a contract between 28 member states whose constitutional ideas and principles form the basis of the European Union. A direct effect of this is Art. 6 (3) of the Treaty of the European Union (TEU). According to this provision the fundamental rights granted in the individual member states' constitution form a part of primary EU law. The second article of the German constitution states that every person has the right to free development of their personality. Similar to that, the Declaration of the Rights of Man and of the Citizen of 1793\footnote{Archives parlementaires, first series, volume VIII, débats du 26 août 1789, p. 489. \href{https://gallica.bnf.fr/ark:/12148/bpt6k495230.image.f557.langFR}{https://gallica.bnf.fr/ark:/12148/bpt6k495230.image.f557.langFR} (last accessed on 2020-08-11).} that forms a part of today France's constitution states that everyone is born and dies as a free man and with the same rights. This shows that also the French constitution is founded on the principle of individuality. Based on this principle of individual freedom, a fundamental right of freedom is acknowledged as a foundation of EU law despite no direct mentioning of it in the EU CFR~\cite[p.244]{zeup2014}. This fundamental freedom right together with the idea of the individual freedom to develop one's personality as it is directly stated in the German constitution stems from the Kantian ideal of freedom that heavily influenced the occidental culture and understanding of the law. According to Kant, every person possesses the primal right of freedom from which they can develop their personality~\cite[p.237]{kant1870grundlegung}. This freedom right to develop one’s own personality, however, cannot comply with fixed quotas as a measure of fairness. Quotas as fairness measures exchange individuality for a seemingly common goal, they limit private autonomy to fulfill a social goal. The discussion on fairness always relies on a trade-off between individuality and commonality. As such, defining fairness only based on an understanding of fixed quotas not based on ground truth is no sufficient compromise between individuality and commonality but a one-sided emphasis on commonality. It violates the right to develop one’s own personality out of one’s individual freedom that is inherent in EU primary right.

This does not change with the ECJ’s fairness measure of \textit{negative dominance}. It is true that this fairness measure only looks at fixed distributions without regarding individual properties. However, it is very important to consider the context of this fairness measure. While discrimination in algorithmic decision making processes cannot be attributed to a neutral rule but only to the input data set, the ECJ always dealt with cases where it knew the underlying neutral rule and could therefore find justification for the less favorable treatment identified after having identified negative dominance. In non-AI decision cases, the ECJ has a powerful correction tool with the possibility to justify less favorable treatment with the neutral rule that lies behind it. This process changes dramatically with the introduction of ADM systems. Without the possibility for the judge to analyze the underlying neutral rule and its purposes a justification is no longer possible. A less favorable treatment means discrimination. As such, the situation must be distinguished from the one described above. The measure of \textit{negative dominance} can no longer be held in the context of algorithmic decision making. Without the means of justification inside the discrimination term, the tool for the judge to guarantee justice in each individual case and by that to guarantee the development of each one’s individual freedom is removed. The negative dominance test that relies on the fairness idea of Independence is therefore insufficient for ADM systems.

\subsubsection{Conditional Independence}
The lack of consideration of individual rights could be solved by a broad modification of the \textbf{Independence} fairness measure. In contrast to \textbf{Independence}, for \textbf{Conditional independence} some properties in the data set are defined as a condition based on which the data set is split into individuals who fulfill the condition and individuals who do not. Then, only for the group that fulfills the defined condition must the result distribution be the same. \textbf{Conditional independence} thus tries to solve the problem of disregarding individual qualities by choosing some of the most deciding factors. Nevertheless, it is not further bound to the ground truth but relies on a simple prediction parity. 

As such, conditional independence is a highly attractive fairness measure. It is the idea of equality of opportunity that only the characteristics of a person are important, which they themselves can change, or at least influence. When choosing such features as the key factor for fairness, \textbf{Conditional independence} corresponds with the constitutional goal of the individual development of one’s freedom. Furthermore, the fact that the ground truth is not regarded, makes the integration of structural imbalances such as racism or sexism at least less likely.

However, serious legal and political concerns remain: The right to determine the conditions that define the group of people for which a certain quota must be fulfilled means great political power. Different conditions have to be chosen for different scenarios and it is often unclear what the \textit{right} conditions are. Rather, the choice of conditions is founded on political beliefs and individual worldviews. The question whether the \textit{right} conditions can be found therefore has no clear answer. The high significance of such a decision would make the parliament the right place to discuss, however, no parliament can make so many individual decisions for each scenario. Rather, parliament shall make general, broad laws that can be worked out by the executive branch. This leads to a high risk of lobbying influence from many interest groups (e.g. NGOs or companies), which will prefer different conditional properties that they consider best for their own interests.

For many scenarios, a good condition will only be found with a huge amount of information for the deciding organs. This could conflict with data protection provisions or more generally the principle of informational self determination.

Moreover, \textbf{Conditional independence} means that only the parity between the group fulfilling the condition is regarded. People that do not fulfill the condition are not relevant for the determination of fairness. It is likely that in hiring scenarios conditions will be bound to individual performance evaluation values such as school grades or work years. Who fulfills such conditions does, however, also depend heavily on a structural imbalance in the society. For example, household income is negatively correlated with school grades - if some groups more often have a lower household income \footnote{\textbf{Statista}, Monatliches persönliches Nettoeinkommen im Jahr 2006 nach jeweiligem Migrationshintergrund, 
30.06.2009, https://de.statista.com/statistik/daten/studie/150623/umfrage/monatliches-persoenliches-einkom-
men-nach-jeweiligem-migrationshintergrund/ (last accessed 25.04.2021).}, the probability for them to have bad school grades increases.~\cite{ferguson2007impact}. This means that for the decisive group of people fulfilling the condition of having at least a certain final grade, only a small part of the minority group could be considered by the actual fairness measure.
\textbf{Conditional Independence} therefore is much less a way of affirmative action than \textbf{Independence} and therefore not sufficient to achieve equal treatment. Though, exactly this means that affirmative action is not an obligatory part of \textbf{Conditional Independence}. If \textbf{Conditional Independence} was to be used to define "less favorable treatment", affirmative action would therefore not be a mandatory part of any discrimination definition pursuant to section 3 GET. As such, the idea of affirmative action as an additional measure as put in Art. 23 EU CFR remains intact. \textbf{Conditional Independence} is thus much more compliant with EU law for ADM systems.

\subsubsection{Separation}
Different to the aforementioned \textbf{Independence} fairness measures, \textbf{Separation} is based on a ground truth, meaning the statistically captured \textit{true} classification of people in the data set. \textbf{Separation} demands an equal matching of predicted classifications with the \textit{true} classification within each group with a protected feature. Ideally, this would mean that reality would be represented exactly in the model, while the representation would be equally good or bad for each protected group. In a hiring context especially the false negative rates, i.e. the rejection for a job despite qualification (according to data record), should ideally be the same for all groups.

At first sight, \textbf{Separation} seems to correspond really well with the ideal of developing one’s own personality. The people who developed their capabilities matching to the job that is advertised and thus have the best qualification for it, will be chosen by the algorithm. Additionally, \textbf{Separation} is able to mitigate the advantaging of the majority group, meaning, the group with numerical dominance of individuals e.g. of an ethnic, that is inherent to most machine learning algorithms.

The dependence of \textbf{Separation} on a ground truth, however, entails a severe problem: First, there can be reasonable doubts whether the captured data always truly represent reality. Data must be available to be collected and will therefore be less likely to capture minority information~\cite{derose2013race}. As such, the information that an machine learning model can gain from the data set for minorities will often be insufficient so that reality cannot be accurately represented by the machine learning model’s prediction. \textbf{Separation} therefore needs a strong regulation of non-discriminatory data sets to prevent a faulty mapping of reality. In contrast to the actual goal of \textbf{Separation}, a discriminatory mapping could reinforce structural social imbalances instead of mitigate them, while still fulfilling the \textbf{Separation} fairness conditions.

Second, conversely to \textbf{Independence} measures, \textbf{Separation} directly integrates structural imbalances of the society into the model’s final decision. As the goal is to perfectly map the decision to the ground truth, structural disadvantages for minority groups will also be mapped to the model's prediction while \textbf{Separation} is still fulfilled~\cite{hardt2016equality}. Looking at Art. 23 EU CFR, EU secondary anti-discrimination law and Art. 157 (4) TFEU the legal goal of mitigating said structural imbalances in society is clearly stated, which seems to conflict with \textbf{Separation}.  However, as already shown in chapter~\ref{chap:independence}, EU law allows for affirmative action in some cases as single additional measure, but not as a broadband solution\footnote{Compare Art. 23 EU CFR}. By contrast, directly integrating affirmative action into the actual discrimination definition is severely flawed. Thus, it seems compliant with EU law to implement \textbf{Separation} into the model as a fairness measure while at the same time taking affirmative actions to mitigate structural imbalances.
To mitigate the risks of \textbf{Separation} reinforcing structural inequalities in society, the debiasing of data sets prior to the training seems promising~\cite{calmon2017optimized}. Lawmakers could therefore put up specific debiasing requirements that must be fulfilled by companies before developing an ADM system. However, it is important to note that perfect debiasing does not (yet) seem possible~\cite{gerritse2020effect}.

\subsubsection{Sufficiency}
Very close to \textbf{Separation}, \textbf{Sufficiency} also demands the mapping of a ground truth to predictions. In fact, \textbf{Sufficiency} just defines a different relation of values to be considered fair compared to \textbf{Separation}. As such, almost all points stated for \textbf{Separation} apply here as well. However, there is a distinctive difference.

Different to \textbf{Separation}, \textbf{Sufficiency} allows for differing false positive rates between protected groups. Regarding the case of COMPAS, a tool that predicted the probability of a person's recidivism, COMPAS’ predictions led to a higher false positive rate for African-Americans, meaning that they were falsely predicted more often to be recidivist, which eventually led to the judge’s decision to detain them~\cite[p.9]{dieterich2016compas}. In such a case the different false positive rate leads to a higher risk for a person belonging to a minority group to be violated in their rights without having any chance to mitigate that risk. This does not comply with the idea of member states’ constitutions and EU primary law that each person shall be able to develop their personality individually, meaning based on their own merits and the traits they can influence. 
\textbf{Sufficiency} can, nevertheless, comply with EU law in situations where false positive rates do not lead to a rights infringement (for example the promotion of a person that is not qualified according to the ground truth).
In such situations, the same caveats described for \textbf{Separation} apply to \textbf{Sufficiency}, in particular the problem of mapping predictions to an inherently racist, sexist or otherwise structurally imbalanced ground truth.

\subsubsection{Individual Fairness}
As the name suggests, \textbf{Individual Fairness} tries to establish fairness by focusing on the individual information in a data set without regarding the group it belongs to. \textbf{Individual Fairness} does not look at groups but at each individual with its unique mix of properties. The fairness measure works under the premise that people who are hired for a certain job and do it well have certain similarities in their characteristics. Thus, for machine learning \textbf{Individual Fairness} means that individuals who have a similar mix of properties get similar prediction outcomes, e.g. classifications.

This aspect of \textbf{Individual Fairness} raises some issues. At first, defining a distance metric that measures the similarity between individual in a data set is a demanding task that cannot be perfectly fulfilled. The more properties there are for an individual in a data set the more dimensions must be connected in one metric. Imagine a person is very similar (close) to another person in three characteristics, but extremely dissimilar (far away) in three other characteristics. The multidimensionality of the characteristics makes it almost impossible to define a perfect metric.

Additionally, it is quite possible that especially people belonging to minority groups will differ from the general property set. This does not necessarily mean that they are less qualified, but rather that they may have an unusual type of expertise that is equally, perhaps even more useful.

\textbf{Individual Fairness} is hardly able to map these qualifications which deviate from the general set of characteristics~\cite[p.13]{friedler2016possibility}. Rather, it heavily advantages the existing majority groups representing the majority of people who will be positively classified. These majority groups form the property set that \textbf{Individual Fairness} compares people with to determine their similarity. 

This means that \textbf{Individual Fairness} actually asks from minority groups to adapt to the property set of the majority group. The ideal of \textbf{Individual Fairness} and of individuality as it is aimed at by EU CFR and German or European constitutions though does not mean the forming of a homogeneous prototype of persons. Each person shall develop their own personality and shall not be pressured to give up their own personal traits. In addition, companies are often not interested in employing mainly employees who are very similar in their skills and characteristics. Instead, employees with so-called "t-shaped skills" are often sought~\cite{spohrer2010integrated}. These are experts who are profound problem solvers in their home discipline, but who are also able to interact with and understand specialists from a wide range of disciplines and functional areas~\cite{barile2013dynamic}.

Even if \textbf{Individual Fairness} pretends only to regard the individual, it in fact leads to a decomposition of individuality in favor of a homogeneous majority driven prototype personality. This consequence is not compliant with the ideal of individuality and development of the individual person’s freedom as laid down in~\ref{chap:independence} and cannot be implemented into a machine learning model in a EU law compliant way.

\subsubsection{Counterfactual Fairness}

The idea of \textbf{Counterfactual Fairness} is to determine whether a decision outcome would be significantly different if a data item belonging to a protected group was not part of a protected group, or vice versa. On the basis of certain assumptions, namely the attributes relevant for a decision and their relationship to each other, coefficients are calculated which attempt to extract the influence of protected features on non protected features and the decision output. 

%Only by that can the counterfactual of a person have the exact same decision output. Counterfactual Fairness thus heavily diverges from the notion of a result-based assessment and assumes that fairness can only be achieved by analyzing and imitating a decision process. 

%The complex relations between protected and unprotected features and the final decision have to be analyzed and determined. In imitating a human decision process, hidden variables that are deemed to lie behind the observable variables such as grade or years of work experience, e.g. knowledge or diligence, are thought of. The computation of coefficients is based on the predictions of a machine learning model. The gained decision output distribution is used to build a causal relationship model. This causal relationship model is then used to create counterfactuals of actual input data points. These counterfactuals can then be fed into the machine learning model to check whether the model's prediction remains the same. Should that not be the case, the operator can train the model with created counterfactuals to let it learn the relationships underlying the causal graph. 

Since getting 100 \% identical results for all counterfactual pairs is highly unlikely, a certain threshold of counterfactual fairness that must be reached can be established. 

%Still changing the decision output distribution, e.g. by adding new training data, can easily require a new computation of the whole causal model if counterfactual fairness shall still be guaranteed. 

This idea resembles \textbf{Conditional Independence} in the fact that a certain set of rules (in the context of counterfactual fairness the causal graph) must be specified. This raises the usual questions regarding an authority that specifies such a set of rules. Although highly problematic from the perspective of governmental competence, the creation of such causal graphs can only be left to domain expert groups that consist of the developer or operator of a model and people knowledgeable in the relationships of certain attributes in the context of the respective application. In hiring scenarios this could be an HR company person and an equal opportunities officer together with a developer of the respective model that knows its exact learning attributes.
%It is only him that knows the exact data used for training and that knows the decision model best. 

Due to the high correlation of the causal graph and the specific decision model a democratically legitimized body could never create a sufficiently precise causal graph that would guarantee counterfactual fairness. The role of the state is rather that of an observer than a decider. This is even more problematic as there is a distinctive difference to \textbf{Conditional Independence}: Whereas the latter leaves the actual decision process to the machine, \textbf{Counterfactual Fairness} goes even beyond that: the whole decision process shall be generally contrived with all its regarded attributes and interdependencies while computing out any dependence on any protected feature. It resembles the task of finding the world formula to establish a perfect causal model that remains fair and correct in its assumptions, does not take any protected feature or any feature that is necessarily depending on it into account, and still retains a sufficient meaningfulness. This task must be left to private domain expert groups that would be foremost bound by the \textbf{Counterfactual Fairness} dogma. In order to still uphold public control and leave the competence of forming society to the state, executive authorities as well as judges could have the right to audit the causal graph to look for glaringly false assumptions. However, judges cannot have the power to question every single aspect and coefficient of the causal graph. The idea of a process-oriented assessment should not mean that the whole process is replaced by the judge's decision process.

It is today generally accepted that a judge cannot fully revise decisions of public independent expert bodies as they enjoy a so called margin of judgment\footnote{See especially for expert bodies: Federal Administrative Court (Germany), decision from 16/12/1971 - I C 31/68 = NJW 1972, 596.}~\cite{bachof1955}. This means that courts can only check for certain decision errors: misjudgment of the existence of a margin, taking as a basis false facts, ignoring generally accepted assessment criteria, making extraneous considerations or exceeding the given margin of judgment. Leaving public authorities with such a margin is especially problematic as they are legally bound to guarantee the fundamental rights granted in the constitution. A domain expert group, by contrast, is a private entity and as such not directly bound to the constitution. This could facilitate a margin of judgment for such expert groups towards jurisprudence. In the end it is left to legislation whether domain expert groups' decisions shall be fully revisable by a judge.
However, it is to note that, since a judge is in principle free in his decisions to enforce the law, this might as well mean that the arbitrariness of a private decision maker in creating the causal graph could be replaced by that of a judge. The existence of arbitrariness in decision processes is eventually a non-deniable and inevitable fact. 

The idea of assessing fairness by regarding the decision process is obviously compliant with current EU law. Still, there remains a significant issue: \textbf{Counterfactual Fairness} upsets the whole legal conception of equality, not in questioning individuality but in showing that our process-based assessment is inherently flawed. When we ask an employer not to regard any protected feature for their decision, we explicitly ignore protected features in a way of fairness through unawareness. \textbf{Counterfactual Fairness}, however, shows that a recognition of such features is also necessary in a process-based assessment to establish a decision, where for example gender indeed has no effect on the decision anymore. Grades that are supposed to carry an objective value actually depend on protected features that cannot or do not have to be changed (e.g. religion). The calculation of an unbiased grade would then mean a different significance of the same grade value in the decision process. Implementing this notion of fairness would leave us with a huge problem: Today's legal approach necessarily perpetuates or at least braces structural imbalances in the society. Considering a grade while being unaware of protected features actually implies a biased decision as grades are in reality not completely objective but influenced by immutable group attributes that are not connected with individual performance~\cite{fischer2013sex}. However, ceasing to do so, the fundamental notion of equality as unawareness in the decision process can no longer be upheld. This poses a highly problematic question: is it compliant with the law to treat objective values differently between groups to reach a decision that is actually unaware of a protected feature as in \textbf{Counterfactual Fairness}? Such a question has never had to be answered as the internal decision making processes of humans have never been made apparent. The idea of anti-discrimination law lies in treating everyone the same unaware of their protected features.\footnote{See Art. 3 section 3 BL: \textit{No person shall be favored or disfavored because of sex, parentage, race, language, homeland and origin, faith or religious or political opinions}.} This means that a decision that is different only based on a different protected feature (e.g. male vs. female) is in principle not compliant with anti-discrimination law. However, one cannot deny that treating objective values differently based on a protected feature also constitutes a less or more favorable treatment, this time only inside the actual decision process. This less or more favorable treatment represents a direct discrimination as the treatment is openly different and not based on a seemingly neutral provision. As such, no general justification of such treatment should be possible. This raises the question whether we humans do also implicitly deal with objective values differently. This is even more evident when one considers that counterfactual unfairness of a machine learning model results from real world data which is based on human decision making. When, in other words, our notion of equality and process-based assessment is inherently flawed and \textbf{Counterfactual Fairness} reaches the goal that anti-discrimination law actually set out, society and lawmakers have to severely ask themselves if current anti-discrimination law has to be amended or at least widely construed so that the goal of a not only formal but substantive equality will be reached.

\subsubsection{Legal Practice}
In practice, several questions arise how fairness measures and fairness assessments can be implemented into anti-discrimination lawsuits. The court is asked to determine if a discrimination is given in a specific context. Pursuant to Section 22 GET, the plaintiff needs to establish facts from which it may be presumed that there has been discrimination based on one of the features protected by the GET so that it will be for the defendant to prove that there has been no breach of the provisions prohibiting discrimination. According to the legislator's will, this means that the plaintiff has to fully prove a less favored treatment but can rely on the provision of circumstantial evidence to prove the causality between the treatment and the plaintiff's protected feature.\footnote{BT-Drs. 16/1780, 47} Requiring the full proof of a less favored treatment by the plaintiff though gets complicated in case of algorithmic decision making. As we have stated, the term "less favored treatment" encompasses itself the question of fairness and as such a legal policy. Without the data that the respective decision model was trained on, no plaintiff will be able to prove such less favored treatment. The law should therefore be construed in such a way that the plaintiff must only prove that it belongs to a disadvantaged group, e. g. has not been hired or has not been given a loan. Consecutively, the plaintiff then must only indicate a link between the disadvantaging decision and their protected feature by circumstantial evidence. If done so, the defendant must then prove to the judge that the decision model used by them is not discriminatory. The judge must be able to cross-check the allegedly discriminating decision with the respective fairness measure that applies to the context wherein the decision has been made. In each case, this means that the court must be provided with data that proves the compliance of the defendant's decision model. In case of fairness measures not dependent on a ground truth, the defendant could provide the court with synthetic, meaning artificially created, data that, fed into the decision model, leads to the group distribution wished for. 

However, already in this case, it could be problematic to rule out a possible discrimination of an individual person based on synthetic data. Who knows whether the model reacts the same way to real world data? As such, it would be favorable for the defendant to provide the court with real data from his decision process to prove total compliance of the decision process in question with the relevant fairness measure. 

In contrast, the verification of fairness measures that depend on a ground truth appears much more difficult. In this case the question arises which data set shall be considered as ground truth. 
A conceivable option would lie in a public balanced data set that entails every group of people in a society and appropriately displays group distributions in a society. However, such a data set would also have to be linked to "true" outcomes for each of the persons in the data set. Determining all "true" outcomes for decisions through governmental instruments is at the same time an immense intrusion of civil freedoms and extremely inefficient and slow. By prescribing "true" values for who for example should get a job and who should not, such a public ground truth pretends to follow allegedly "correct" societal ideals of a society. If such a ground truth is adapted and implemented by the state, it becomes an enforceable imperative. Any differing ideas of society would have to be discussed through state representatives but could not anymore be realized by one's own decisions. This is essentially the death of a liberal civil and decentralized society.

If there is no such public ground truth, the ground truth for the court to base its judgment on can only consist of the data set that was to some degree used to train the respective model which is allegedly discriminatory. This poses the question which data sets must be provided to the court. Assuming that the provider has not trained the model by its own, the actually decisive data set is the training data set that was used by the model developer. When the provider was not additionally training the model, it is operated with the same ground truth used by the developer in the training process.
Without provision of a data set that serves as the ground truth, no judge can determine if a fairness measure dependent on a ground truth like Separation and Sufficiency is fulfilled.
The lack of a ground truth provided could therefore lead to a loss of the law suit by the defendant. The defendant must provide the judge with the correct ground truth data. 
In case of an anti-discrimination proceeding litigated before a German court, the defendant can request expert evidence pursuant to section 402 seqq. German Code of Civil Procedure (CCP) to prove whether the algorithm used by the defendant is discriminatory or not. However, if the ground truth required for that check is in the hands of a third party, this party cannot be forced to make their data available to the expert. The defendant can try to let the third party provide the data required by summoning them as a third party in the legal proceedings pursuant to section 72 CCP. As a consequence, according to section 68 CCP, the third party will not be heard with the objection that the previous law suit was falsely decided when the defendant will sue them in order to take recourse. The so called third-party notice is a general instrument in EU countries as it is acknowledged as a procedural option in Art. 65 Brussels I-Regulation\footnote{As civil procedure legislation is a national competence, Art. 65 Brussels I-Regulation does not regulate an EU-wide third-party notice instrument itself. Rather, it decides which civil procedure legislation applies in case of multi national cases. As the Brussels-I Regulation refers to the option of third-party notice, one can assume that every EU member state has such an option in its respective civil procedure legislation in some form.}

Filing a third-party notice could therefore let the third party provide the data required to prevent a negative outcome of the law suit in order to prevent the following recourse. However, the third party is in no way obliged to join the proceedings and provide the ground truth data. If they deem their information advantage so valuable that they do not want to share it in any way, the civil proceedings legislation allows them to lose the law suit against them without having to make any of their ground truth data public. In that way, a third party such as the developer of an allegedly discriminatory algorithm, could prevent a thorough analysis of its algorithm. The so called principle of production of evidence means that it is always up to the parties of a legal proceeding to produce evidence. This principle leads to the consequence that civil legal proceedings are not sufficient to guarantee discrimination-free algorithms that comply with fairness measures. Therefore strong auditing rights of public authorities and a sufficiently strong competition law are necessary to effectively establish fairness measures.

\section{Discussion and future work}
\label{section:discussion}

The subject of fairness in algorithmic decision making is a highly significant issue that poses complex questions for the law and questions several of its fundamental notions.
We argued that \textbf{Conditional Independence}, \textbf{Separation}, \textbf{Sufficiency} and \textbf{Counterfactual Fairness} can all possibly be implemented in compliance with EU law dependent on their concrete application. All four of them follow the idea that fairness and equality or what defines “a less favorable treatment” can be expressed in a formal mathematical function.\footnote{See especially \textit{On the (im)possibility of fairness} \cite{friedler2016possibility} for a strong argumentation towards a formal expression of fairness.} This mathematical focus leads to a result-oriented understanding of equality, that prioritizes output distributions, which interferes with the notion of individual freedom. Only \textbf{Counterfactual Fairness} is based on a process-oriented understanding of equality that is then connected with a result-mapping process. The idea of a counterfactually fair decision output is fascinating as it simultaneously corresponds with the goal of anti-discrimination law and infringes it in its process. Therefore, it remains to be seen whether and how exactly \textbf{Counterfactual Fairness} can be implemented in an EU law compliant way. However, in our opinion lawmakers should further investigate and elaborate the idea of \textbf{Counterfactual Fairness} as it keeps the ideal of individual freedom and personality in algorithmic decision making and thus prevents a society that is shaped through endless quotas that are centrally determined by governmental bodies. It is the core of a free civil society that the people can make decisions based on their own preferences which differ one from another.

%Therefore, it is important to note that a fairness measure is supposed to be the basis on which the entire discrimination term for AI decisions will be founded

EU law recognizes the need of affirmative action and result driven quotas and explicitly allows them. Though, it also assumes that such action shall not be the core of the notion of equality but rather one of several means to establish equality in society. The more only a group centered result accounts for anti-discrimination legislation, the less the individual in the group is accounted for. In the end, EU Law and especially the EU CFR only determine the outer boundaries of the trade-off between group focus and individual focus, between result focus and process focus. It remains a hugely political question whether fairness measures should be broadly implemented and if so what fairness is preferable for our society. Lawmakers have to decide together with politicians and domain experts which kind of fairness is applicable under which condition~\cite{bernstein2018}. To do so, they need to be thoroughly informed what kind of fairness has which kind of implications and consequences on short and long term. 

Many fairness measures are mutually exclusive (e.g. independence, separation, and sufficiency are mutually incompatible with each other~\cite{barocas2018fairness}). Due to the different interests and interest groups in different decisions, for example in deciding on a job hire in contrast to the decision of receiving a loan, different fairness measures fit better to one than to another decision situation. It should be noted that \textbf{Counterfactual Fairness} of the fairness measures analyzed in this paper would in practice give decision-makers the greatest room of freedom to determine the decision conditions themselves while state bodies, such as courts and supervisory authorities, would remain monitoring bodies. What fairness measure should be applied to what decision situation, should be decided by a democratically legitimized organ, ideally the parliament. The respective organ could work out general interests in situations that point to one or another fairness measure. However, there is an immense amount of decisions and varying interests even in the same sector, e.g. in HR, where the decision on who is promoted can be driven by different interests than the decision on who is hired at all. Finding the best fitting measure for each of these decisions without overly generalizing is an enormous task. The term \textit{less favorable treatment} is indeed not defined with good reason: its notion is highly context-dependent and a less favorable treatment in one case does not necessarily need to be one in another. This makes it impossible to find a general definition for a less favorable treatment by one of the fairness measures that applies to every case. Here, again, \textbf{Counterfactual Fairness}, which allows decision makers to follow their own causal graph, could be a solution that does not require a state organ to establish perfect abstract general rules.

If fairness measures are to be used to evaluate the appropriateness of AI based ADM systems, it must also be taken into account that many software products are regularly updated. Many AI systems can even continue to learn in use and thus, at least theoretically, change their behavior at any time. It is therefore necessary to at least consider the use of fairness measures for a continuous evaluation system. After all, the fairness requirements should be met throughout the entire life cycle of an AI product. There must be an escalation/emergency procedure to correct unforeseen cases of unfairness when they are discovered. The enormous potential impact of ADM systems on societies makes a mechanism to continuously review fairness desirable. Instead of legislating that fairness measures should be implemented directly into AI models, we suggest using fairness measures rather as an indicator that allows for a result-oriented review of a decision output of a particular AI model. This could be done either by a supervisory authority in the case of examinations and audits or by judges in the case of discrimination claims. Furthermore, the question whether proposals from an AI model should be put into practice while there exist unresolved problems of structural racism and sexism in a society, should have a significant influence on the further discussion. Without specific measures, AI tends to learn and therefore to perpetuate or even to propagate structural inequalities in society through its learning heuristics~\cite[p.2]{barocas2016big}. An answer to the question of whether it must really be the task of an AI system to counteract existing inequalities, e.g. by being forced to adjust mathematical formulas, or whether this is a social task which constitutes the necessary basis for the use of such systems, seems to be essential to get closer to answers to many legal questions and problems. 

The process-oriented approach in current anti-discrimination legislation raises the question whether fairness can and should be mathematically operationalized at all. The mathematical operationalization of fairness leads to a result driven anti-discrimination law in which seemingly individual justice is preprogrammed and predetermined without many possibilities for the individual to contest this predetermination~\cite{binns2018s}. The fact that the law uses the term indirect discrimination shows that it is aware of the phenomenon of statistical discrimination. However, the consideration of indirect discrimination is still only one part of anti-discrimination legislation and has not yet led to a fundamental change in legislation towards a result-oriented assessment of discrimination. This is possibly bound to change with algorithmic decision making systems which often do not allow for the transparency needed in their decision making processes and thus make a control of decision outputs inevitable. This raises the question whether a broad result control is compliant with EU law.

Due to the fact that result driven fairness measures are a problem, developers should always try to create models in a way that makes the decision process transparent. If possible, a whitebox model should always be used to increase the transparency and explainability of an AI-based algorithmic decision system. It is often argued that blackbox systems, such as artificial neural networks, are clearly superior to whitebox systems in terms of precision, significance and efficiency~\cite{dunis2002modelling, zekic2004small, wang2009comparison}. At the same time, the technical literature also shows in some examples that this is not always the case~\cite{rudin2019stop}. We see a possible approach to give weight to these considerations in the obligation for those responsible (whereby the question of responsibility must first be clarified) to provide objective and, in the best case, verifiable reasons why no whitebox model could be used. A whitebox model makes it possible to analyze the model's decision process, not only its found results. By that, a process-oriented assessment can be upheld and the competence of judges to establish individual justice to people allegedly discriminated is preserved.\footnote{See also German Federal Constitutional Court, judgment from 19 May 2020 - 1 BvR 2835/17, notice 192.} On top of that, the explainability of AI guards a human-centered understanding of society in which the control of processes and results is always in human hands. 

Additionally, AI poses a challenge to the entire scope of non-discrimination law itself. Fairness measures may relate to any particular group, which may be at an advantage or disadvantage in the context in which the system is being deployed. It cannot be assumed that disparity only occur between legally protected groups. Groups which do not map to legally protected characteristics may suffer levels of disparity which would otherwise be considered discriminatory if applied to a protected group~\cite[p.11]{wachter2020fairness}. To put it bluntly, all people who have a birthmark under their right eye and have never been in a sports club for more than three years could be discriminated against for example. This raises the question whether such discrimination must also be prevented. Specific rights of equality resemble liberty rights in that they provide specific protection comparable to that of freedom rights\footnote{See Baer\&Markard, in~\cite{grundgesetz2018} to Art. 3 BL, recital 404}. 

In contrast, the general right of equality (all people are equal under the law) does not contain any qualitative conditions but simply consists of a comparison between any two or more groups. If now any distinct group discriminated by an algorithmic decision was to be protected with a specific equality right, the border between general and specific equality rights becomes blurred. Therefore, such an extension is at least debatable.

Algorithms and fairness measures yield a great potential for finding/exposing problems. In case of specific suspicion though we think a thorough examination by hand needs to be done despite the results of fairness measures.

%Optional to-dos:
%- we have 3 arxiv sources, maybe we find better ones (currently 17,50 and 53)
\section{Acknowledgement}

The research was performed within the project GOAL \enquote{Governance of and by algorithms} (Funding code 01IS19020; \href{https://goal-projekt.de/en/}{https://goal-projekt.de/en/}) which is funded by the German Federal Ministry of Education and Research. The content of this paper is the sole responsibility of its authors.

Moreover, the research  has also been  conducted  within  the  project  ”Fairand Good ADM” (16ITA203) funded by the Federal Ministryof Education and Research (BMBF) of Germany.



%% References


\bibliographystyle{elsarticle-num}
\bibliography{literatur}

\begin{thebibliography}{10}
\expandafter\ifx\csname url\endcsname\relax
  \def\url#1{\texttt{#1}}\fi
\expandafter\ifx\csname urlprefix\endcsname\relax\def\urlprefix{URL }\fi
\expandafter\ifx\csname href\endcsname\relax
  \def\href#1#2{#2} \def\path#1{#1}\fi

\bibitem{altmann2017autonomous}
J.~Altmann, F.~Sauer, Autonomous weapon systems and strategic stability,
  Survival 59~(5) (2017) 117--142.

\bibitem{shapiro2017reform}
A.~Shapiro, Reform predictive policing, Nature News 541~(7638) (2017) 458.

\bibitem{narula2018everyday}
G.~Narula, Everyday examples of artificial intelligence and machine learning,
  Tech Emergence (2018).

\bibitem{muller2020impact}
C.~Muller, {The Impact of Artificial Intelligence onHuman Rights, Democracy and
  the Rule of Law}, Tech. rep., Council of Europe (2020).

\bibitem{ameriks2000aristotle}
K.~Ameriks, D.~M. Clarke, Aristotle: Nicomachean Ethics, Cambridge University
  Press, 2000.

\bibitem{rawls1971theory}
J.~Rawls, A theory of justice, Harvard university press, 1971.

\bibitem{dworkin1981equality}
R.~Dworkin, What is equality? part 1: Equality of welfare, Philosophy \& public
  affairs (1981) 185--246.

\bibitem{dworkin1981equality2}
R.~Dworkin, What is equality? part 2: Equality of resources, Philosophy \&
  public affairs (1981) 283--345.

\bibitem{gurouglu2010unfair}
B.~G{\"u}ro{\u{g}}lu, W.~van~den Bos, S.~A. Rombouts, E.~A. Crone, Unfair? it
  depends: neural correlates of fairness in social context, Social Cognitive
  and Affective Neuroscience 5~(4) (2010) 414--423.

\bibitem{zweig2018fairness}
K.~A. Zweig, T.~D. Krafft, Fairness und qualit{\"a}t algorithmischer
  entscheidungen (2018).

\bibitem{dattner2019legal}
B.~Dattner, T.~Chamorro-Premuzic, R.~Buchband, L.~Schettler, The legal and
  ethical implications of using ai in hiring, Harvard Business Review 25
  (2019).

\bibitem{molnar2018guide}
C.~Molnar, A guide for making black box models explainable, URL:
  https://christophm. github. io/interpretable-ml-book, Accessed 26.08.2020
  (2018).

\bibitem{sokol2020explainability}
K.~Sokol, P.~Flach, Explainability fact sheets: a framework for systematic
  assessment of explainable approaches, in: Proceedings of the 2020 Conference
  on Fairness, Accountability, and Transparency, 2020, pp. 56--67.

\bibitem{shapley1953value}
L.~S. Shapley, A value for n-person games, Contributions to the Theory of Games
  2~(28) (1953) 307--317.

\bibitem{sundararajan2019many}
M.~Sundararajan, A.~Najmi, The many shapley values for model explanation, arXiv
  preprint arXiv:1908.08474 (2019).

\bibitem{ribeiro2016should}
M.~T. Ribeiro, S.~Singh, C.~Guestrin, "why should i trust you?" explaining the
  predictions of any classifier, in: Proceedings of the 22nd ACM SIGKDD
  international conference on knowledge discovery and data mining, 2016, pp.
  1135--1144.

\bibitem{ribeiro2018anchors}
M.~T. Ribeiro, S.~Singh, C.~Guestrin, Anchors: High-precision model-agnostic
  explanations, in: Thirty-Second AAAI Conference on Artificial Intelligence,
  2018.

\bibitem{craven1996extracting}
M.~Craven, J.~W. Shavlik, Extracting tree-structured representations of trained
  networks, in: Advances in neural information processing systems, 1996, pp.
  24--30.

\bibitem{datenschutzrecht2019}
S.~Simitis, G.~H. (Herausgeber), I.~S. gen. Döhmann, Datenschutzrecht, Nomos,
  2019.

\bibitem{bernstein2018}
J.~Bernstein, S.~Elsayed-Ali, E.~Kochi, M.~Patel, How to prevent discriminatory
  outcomes in machine learning, Tech. rep., World Economic Forum (2018).

\bibitem{guggenberger2019}
N.~Guggenberger, Multimedia und Recht (MMR), C.H. Beck, 2019, pp. 777--778.

\bibitem{verma2018fairness}
S.~Verma, J.~Rubin, Fairness definitions explained, in: 2018 IEEE/ACM
  International Workshop on Software Fairness (FairWare), IEEE, 2018, pp. 1--7.

\bibitem{hutchinson201950}
B.~Hutchinson, M.~Mitchell, 50 years of test (un) fairness: Lessons for machine
  learning, in: Proceedings of the Conference on Fairness, Accountability, and
  Transparency, 2019, pp. 49--58.

\bibitem{friedler2019comparative}
S.~A. Friedler, C.~Scheidegger, S.~Venkatasubramanian, S.~Choudhary, E.~P.
  Hamilton, D.~Roth, A comparative study of fairness-enhancing interventions in
  machine learning, in: Proceedings of the conference on fairness,
  accountability, and transparency, 2019, pp. 329--338.

\bibitem{barocas2018fairness}
S.~Barocas, M.~Hardt, A.~Narayanan, Fairness and Machine Learning,
  fairmlbook.org, 2019, \url{http://www.fairmlbook.org}.

\bibitem{corbett2017algorithmic}
S.~Corbett-Davies, E.~Pierson, A.~Feller, S.~Goel, A.~Huq, Algorithmic decision
  making and the cost of fairness, in: Proceedings of the 23rd ACM SIGKDD
  International Conference on Knowledge Discovery and Data Mining, ACM, 2017,
  pp. 797--806.

\bibitem{berk2018fairness}
R.~Berk, H.~Heidari, S.~Jabbari, M.~Kearns, A.~Roth, Fairness in criminal
  justice risk assessments: The state of the art, Sociological Methods \&
  Research (2018) 0049124118782533.

\bibitem{chouldechova2017fair}
A.~Chouldechova, Fair prediction with disparate impact: A study of bias in
  recidivism prediction instruments, Big data 5~(2) (2017) 153--163.

\bibitem{kusner2017counterfactual}
M.~J. Kusner, J.~Loftus, C.~Russell, R.~Silva, Counterfactual fairness, in:
  Advances in neural information processing systems, 2017, pp. 4066--4076.

\bibitem{kilbertus2017avoiding}
N.~Kilbertus, M.~R. Carulla, G.~Parascandolo, M.~Hardt, D.~Janzing,
  B.~Sch{\"o}lkopf, Avoiding discrimination through causal reasoning, in:
  Advances in Neural Information Processing Systems, 2017, pp. 656--666.

\bibitem{singh2018fairness}
A.~Singh, T.~Joachims, Fairness of exposure in rankings, in: Proceedings of the
  24th ACM SIGKDD International Conference on Knowledge Discovery \& Data
  Mining, 2018, pp. 2219--2228.

\bibitem{beutel2019fairness}
A.~Beutel, J.~Chen, T.~Doshi, H.~Qian, L.~Wei, Y.~Wu, L.~Heldt, Z.~Zhao,
  L.~Hong, E.~H. Chi, et~al., Fairness in recommendation ranking through
  pairwise comparisons, in: Proceedings of the 25th ACM SIGKDD International
  Conference on Knowledge Discovery \& Data Mining, 2019, pp. 2212--2220.

\bibitem{dwork2012fairness}
C.~Dwork, M.~Hardt, T.~Pitassi, O.~Reingold, R.~Zemel, Fairness through
  awareness, in: Proceedings of the 3rd innovations in theoretical computer
  science conference, ACM, 2012, pp. 214--226.

\bibitem{hardt2016equality}
M.~Hardt, E.~Price, N.~Srebro, et~al., Equality of opportunity in supervised
  learning, in: Advances in neural information processing systems, 2016, pp.
  3315--3323.

\bibitem{mcnamara2019equalized}
D.~McNamara, Equalized odds implies partially equalized outcomes under
  realistic assumptions, in: Proceedings of the 2019 AAAI/ACM Conference on AI,
  Ethics, and Society, 2019, pp. 313--320.

\bibitem{barocas2016big}
S.~Barocas, A.~D. Selbst, Big data's disparate impact, Calif. L. Rev. 104
  (2016) 671.

\bibitem{makhlouf2020applicability}
K.~Makhlouf, S.~Zhioua, C.~Palamidessi, On the applicability of ml fairness
  notions, arXiv preprint arXiv:2006.16745 (2020).

\bibitem{russell2017worlds}
C.~Russell, M.~J. Kusner, J.~Loftus, R.~Silva, When worlds collide: integrating
  different counterfactual assumptions in fairness, in: Advances in Neural
  Information Processing Systems, 2017, pp. 6414--6423.

\bibitem{dawid2000causal}
A.~P. Dawid, Causal inference without counterfactuals, Journal of the American
  statistical Association 95~(450) (2000) 407--424.

\bibitem{pearl2000causality}
J.~Pearl, Causality: Models, reasoning and inference cambridge university
  press, Cambridge, MA, USA, 9 (2000) 10--11.

\bibitem{wachter2020fairness}
S.~Wachter, B.~Mittelstadt, C.~Russell, Why fairness cannot be automated:
  Bridging the gap between eu non-discrimination law and ai, Available at SSRN
  (2020).

\bibitem{beckogk2020}
D.~Looschelders, Beck'scher Online- Großkommentar (BeckOGK, C.H. Beck, 2020.

\bibitem{wong2019democratizing}
P.-H. Wong, Democratizing algorithmic fairness, Philosophy \& Technology (2019)
  1--20.

\bibitem{zuiderveen2019algorithmic}
F.~Zuiderveen~Borgesius, Algorithmic decision-making, price discrimination, and
  european non-discrimination law, European Business Law Review (Forthcoming)
  (2019).

\bibitem{fba2011discrimination}
S.~F. FBA, Discrimination law, Oxford University Press, 2011.

\bibitem{wexler2020knowledge}
N.~Wexler, The Knowledge Gap: The Hidden Cause of America's Broken Education
  System--And How to Fix It, Avery, 2020.

\bibitem{zeup2014}
C.~Herresthal, Grundrechecharta und privatrecht - die bedeutung der charta der
  grundrechte für das europäische und das nationale privatrecht, Zeitschrift
  für europäisches Privaterecht ZEuP (2014) 238--280.

\bibitem{kant1870grundlegung}
I.~Kant, Grundlegung zur metaphysik der sitten, Vol.~28, L. Heimann, 1870.

\bibitem{ferguson2007impact}
H.~B. Ferguson, S.~Bovaird, M.~P. Mueller, The impact of poverty on educational
  outcomes for children, Paediatrics \& child health 12~(8) (2007) 701--706.

\bibitem{derose2013race}
S.~F. Derose, R.~Contreras, K.~J. Coleman, C.~Koebnick, S.~J. Jacobsen, Race
  and ethnicity data quality and imputation using us census data in an
  integrated health system: the kaiser permanente southern california
  experience, Medical Care Research and Review 70~(3) (2013) 330--345.

\bibitem{calmon2017optimized}
F.~P. Calmon, D.~Wei, K.~N. Ramamurthy, K.~R. Varshney, Optimized data
  pre-processing for discrimination prevention, arXiv preprint arXiv:1704.03354
  (2017).

\bibitem{gerritse2020effect}
E.~J. Gerritse, A.~P. de~Vries, Effect of debiasing on information retrieval,
  in: International Workshop on Algorithmic Bias in Search and Recommendation,
  Springer, 2020, pp. 35--42.

\bibitem{dieterich2016compas}
W.~Dieterich, C.~Mendoza, T.~Brennan, Compas risk scales: Demonstrating
  accuracy equity and predictive parity, Northpointe Inc (2016).

\bibitem{friedler2016possibility}
S.~A. Friedler, C.~Scheidegger, S.~Venkatasubramanian, On the (im) possibility
  of fairness, arXiv preprint arXiv:1609.07236 (2016).

\bibitem{spohrer2010integrated}
J.~Spohrer, G.~M. Golinelli, P.~Piciocchi, C.~Bassano, An integrated ss-vsa
  analysis of changing job roles, Service Science 2~(1-2) (2010) 1--20.

\bibitem{barile2013dynamic}
S.~Barile, M.~Saviano, Dynamic capabilities and t-shaped knowledge: a viable
  systems approach, Contributions to Theoreticalo and Practical Advances in
  Management. A Viable Systems Approach (VSA), ARACNE Editrice Srl, Roma (2013)
  39--59.

\bibitem{bachof1955}
O.~Bachof, Jursitenzeitung (JZ), Mohr Siebeck Verlag, 1955, p. 97–102.

\bibitem{fischer2013sex}
F.~Fischer, J.~Schult, B.~Hell, Sex differences in secondary school success:
  why female students perform better, European journal of psychology of
  education 28~(2) (2013) 529--543.

\bibitem{binns2018s}
R.~Binns, M.~Van~Kleek, M.~Veale, U.~Lyngs, J.~Zhao, N.~Shadbolt, 'it's
  reducing a human being to a percentage' perceptions of justice in algorithmic
  decisions, in: Proceedings of the 2018 Chi conference on human factors in
  computing systems, 2018, pp. 1--14.

\bibitem{dunis2002modelling}
C.~Dunis, M.~Williams, Modelling and trading the eur/usd exchange rate: Do
  neural network models perform better?, Derivatives use, trading and
  regulation 8~(3) (2002) 211--239.

\bibitem{zekic2004small}
M.~Zekic-Susac, N.~Sarlija, M.~Bensic, Small business credit scoring: a
  comparison of logistic regression, neural network, and decision tree models,
  in: 26th International Conference on Information Technology Interfaces,
  2004., IEEE, 2004, pp. 265--270.

\bibitem{wang2009comparison}
J.~Wang, M.~Li, Y.-t. Hu, Y.~Zhu, Comparison of hospital charge prediction
  models for gastric cancer patients: neural network vs. decision tree models,
  BMC health services research 9~(1) (2009) 161.

\bibitem{rudin2019stop}
C.~Rudin, Stop explaining black box machine learning models for high stakes
  decisions and use interpretable models instead, Nature Machine Intelligence
  1~(5) (2019) 206--215.

\bibitem{grundgesetz2018}
H.~von Mangoldt, F.~Klein, C.~Starck, Kommentar zum Grundgesetz: GG, 7th
  Edition, C.H. Beck, 2018.

\end{thebibliography}






% \end{thebibliography}

\newpage
\section*{Appendix}

\subsection*{Quality Measure}

Here in this section, we explain the performance (quality) measures used in some fairness measures. These quality measures are based on the confusion matrix. Thus, at first, we explain this matrix briefly.

 \textbf{Confusion matrix}
 The confusion matrix (see table \ref{tab:confusion_matrix}) is a 2$\times$2 table that depicts four quality measures, namely the number of true positive, true negative, false-negative and false-positive cases. Various other fairness measures can be derived from it.

 \begin{table}[h]
\begin{tabular}{llllll}
\cline{3-4}

                                                                                                                                 & \multicolumn{1}{l|}{}                                 & \multicolumn{2}{c|}{\cellcolor[HTML]{C0C0C0}Predicted Label}                                                                                                                                                                                                 &                                                                                                                                                     &  \\ \cline{3-5}
                                                                                                                                 & \multicolumn{1}{l|}{}                                 & \multicolumn{1}{l|}{\cellcolor[HTML]{C0C0C0}Positive}                                                                     & \multicolumn{1}{l|}{\cellcolor[HTML]{C0C0C0}Negative}                                                                            & \multicolumn{1}{l|}{\cellcolor[HTML]{E0E1C5}\begin{tabular}[c]{@{}l@{}}   \\   \end{tabular}}                        &  \\ \cline{1-5}
\multicolumn{1}{|c|}{\cellcolor[HTML]{C0C0C0}}                                                                                   & \multicolumn{1}{l|}{\cellcolor[HTML]{C0C0C0}Positive} & \multicolumn{1}{l|}{{\color[HTML]{036400} \begin{tabular}[c]{@{}l@{}}True Positive \\  (TP)\end{tabular}}}                & \multicolumn{1}{l|}{{\color[HTML]{CB0000} \begin{tabular}[c]{@{}l@{}}False Negative \\ (FN)\end{tabular}}}                       & \multicolumn{1}{l|}{\cellcolor[HTML]{C2D9DA}\begin{tabular}[c]{@{}l@{}}True Positive Rate\\  $\frac{TP}{TP+FN}$\end{tabular}}                       &  \\ \cline{2-5}
\multicolumn{1}{|c|}{{\cellcolor[HTML]{C0C0C0}\begin{tabular}[c]{@{}c@{}}Actual (Target)\\  Label\end{tabular}}} & \multicolumn{1}{l|}{\cellcolor[HTML]{C0C0C0}Negative} & \multicolumn{1}{l|}{{\color[HTML]{CB0000} \begin{tabular}[c]{@{}l@{}}False Positive \\ (FP)\end{tabular}}}                & \multicolumn{1}{l|}{{\color[HTML]{036400} \begin{tabular}[c]{@{}l@{}}True Neagtive\\ (TN)\end{tabular}}}                         & \multicolumn{1}{l|}{\cellcolor[HTML]{E7D2BE}\begin{tabular}[c]{@{}l@{}}False Positive Rate \\ $\frac{FP}{FP+TN}$\end{tabular}}                      &  \\ \cline{1-5}
                                                                                                                                 & \multicolumn{1}{l|}{}                                 & \multicolumn{1}{l|}{\cellcolor[HTML]{DAE8FC}\begin{tabular}[c]{@{}l@{}}Precision \\ $\frac{TP}{TP+FP}$\\ \end{tabular}} & \multicolumn{1}{l|}{\cellcolor[HTML]{CDE0CE}\begin{tabular}[c]{@{}l@{}}False Omission Rate \\ $\frac{FN}{FN+TN}$\end{tabular}} & \multicolumn{1}{l|}{\cellcolor[HTML]{CCCAF8}{\color[HTML]{343434} \begin{tabular}[c]{@{}l@{}}Accuracy \\ $\frac{TP+TN}{TP+TN+FP+FN}$\end{tabular}}} &  \\ \cline{3-5}
                                                                                                                                 &                                                       &                                                                                                                           &                                                                                                                                  &                                                                                                                                                     & 
\end{tabular}
\caption{Quality measures for evaluating ADMs  with a binary classification core. }
\label{tab:confusion_matrix} 
\end{table}

\begin{itemize}

\item \textbf{Accuracy}\\
The accuracy is one of the most popular measures for evaluating the performance of binary classifiers. It is the proportion of correctly classified data instances to the number of all data instances. 
\begin{equation}
Accuracy := \frac{TP + TN}{ TP + FP + TN + FN} 
\end{equation}

\item \textbf{True Positive Rate}\\
True Positive Rate (TPR) indicates the proportion of data instances correctly assigned to the positive class to data instances that actually belong to the positive class. 

\begin{equation}
    TPR:=\frac{TP}{TP+FN}
\end{equation}

 \item \textbf{True Negative Rate}\\
True Negative Rate TNR indicates the proportion of data instances correctly assigned to the negative class to data instances that actually belong to class negative.

\begin{equation}
    TNR:=\frac{TN}{FP+TN}
\end{equation}

\item \textbf{False Positive Rate}\\
The False Positive Rate (FPR) is defined as the proportion of data instances of class negative mistakenly assigned to class positive to data instances that actually be in the negative class.\\

\begin{equation}
    FPR:=\frac{FP}{FP+TN} = 1-TNR
\end{equation}

\item \textbf{Precision}\\
  Precision is another measure that computes the fraction of data instances of positive class, which are correctly classified as positive to all data instances assigned by the system to class positive. 
     
     \begin{equation}
    Precision := \frac{TP}{TP+FP} 
\end{equation}

\item \textbf{False Omission Rate }\\
computes the fraction of data instances assigned to negative label that are actually from the positive class.  
\begin{equation}
    FOR := \frac{FN}{FN+TN} 
\end{equation}

\end{itemize}

\end{document}